\documentclass[12pt,a4paper]{article}
\nonstopmode
\usepackage{epsfig}
\usepackage[colorlinks=true]{hyperref}
\usepackage[a4paper]{geometry}
\usepackage{float}
\usepackage[normalem]{ulem}
\usepackage[stable]{footmisc}
\usepackage[utf8]{inputenc}
\usepackage{a4wide}
\usepackage{color}
\usepackage[dvipsnames]{xcolor}
\definecolor{dark-gray}{gray}{0.13}
\usepackage{amsmath}
\usepackage{amsthm} 
\usepackage{amsfonts,mathrsfs}
\usepackage{amssymb}
\usepackage{placeins}
\usepackage[toc,page]{appendix}
\usepackage[ruled,vlined,linesnumbered]{algorithm2e}
\hypersetup{urlcolor=dark-gray,linkcolor=dark-gray,citecolor=dark-gray}
\usepackage[numbers, square, comma, sort&compress]{natbib}

\begin{document}	
\pagestyle{plain}

\makeatletter
\@addtoreset{equation}{section}
\makeatother
\renewcommand{\theequation}{\thesection.\arabic{equation}}
\pagestyle{empty}

\begin{center}
\phantom{a}\\
\vspace{0.8cm}
\scalebox{0.90}[0.90]{{\fontsize{24}{30} \bf{F-theory Axiverse
}}}\\
\end{center}

\vspace{0.4cm}
\begin{center}
\scalebox{0.95}[0.95]{{\fontsize{15}{30}\selectfont   
Sebastian Vander Ploeg Fallon,$^{a}$ James Halverson,$^{b,c}$}}\\
\vspace{0.1cm}
\scalebox{0.95}[0.95]{{\fontsize{15}{30}\selectfont  
Liam McAllister,$^{a}$ and Yunhao Zhu$^{b}$}}
\end{center}

\begin{center}
\vspace{0.25 cm}

\textsl{$^{a}$Department of Physics, Cornell University, Ithaca, NY 14853, USA}\\
\vspace{0.05cm}
\textsl{$^{b}$Department of Physics, Northeastern University, Boston, MA 02115, USA}\\
\vspace{0.05cm}
\textsl{$^{c}$The NSF AI Institute for Artificial Intelligence and Fundamental Interactions}\\

	 \vspace{1.1cm}
	\normalsize{\bf Abstract} \\[8mm]
\end{center}
\begin{center}
\begin{minipage}[h]{15.0cm}

\medskip
We compute the couplings of Ramond-Ramond four-form axions in three  
ensembles of F-theory compactifications, with up to $181{,}200$ axions.  We work in the stretched K\"ahler cone, where $\alpha'$ corrections are plausibly controlled, and we use couplings to certain non-Abelian sectors as a proxy for couplings to photons. The axion masses, decay constants, and couplings to gauge sectors show striking universality across the ensembles. In particular, the axion-photon couplings grow with $h^{1,1}$, and models in our ensemble with $h^{1,1} \gtrsim 10{,}000$ axions are in tension with helioscope constraints. Moreover, under mild assumptions about charged matter beyond the Standard Model, theories with $h^{1,1} \gtrsim 5{,}000$ are in tension with Chandra measurements of X-ray spectra. This work is a first step toward understanding the phenomenology of quantum gravity theories with thousands of axions.
 
\end{minipage}
\end{center}
\vfill
\today 
\newpage 

\setcounter{page}{1}
\pagestyle{plain}
\renewcommand{\thefootnote}{\arabic{footnote}}
\setcounter{footnote}{0}
%
%
\setcounter{tocdepth}{1}

\tableofcontents

\newpage 

\section{Introduction}

Axions are among the most compelling candidates for new particles beyond the Standard Model. 
The QCD axion provides a dynamical solution to the strong CP problem, and one or more axions could make up the dark matter.  Moreover, axions are abundant in string theory, and in many cases are protected by symmetries that allow them to be exponentially lighter than the string scale and the Kaluza-Klein scale.  Thus, string theory axions could
provide a link between quantum gravity and low-energy experiments.

The fact that string compactifications to four dimensions generally include axions was already well-understood in the 1980s, and the phenomenology of many-axion theories, the \emph{string axiverse} \cite{Arvanitaki:2009fg}, has been the focus of much theoretical work in recent years: see e.g.~\cite{Cicoli:2012sz,Halverson:2019kna,Halverson:2019cmy,Agrawal:2019lkr,
Broeckel:2021dpz,Cyncynates:2021xzw,
Demirtas:2021gsq,glimmers,Gendler:2024adn,fuzzy,Reig:2025dqb,Leedom:2025mlr,Petrossian-Byrne:2025mto,Martucci:2024trp,Reece:2025thc,Benabou:2025kgx,Dessert:2025yvk}.
Advances in computational geometry have made it possible to construct ensembles of compactifications and study the resulting axion physics.  Such analyses have been carried out in considerable detail for axions descending from the Ramond-Ramond four-form $C_4$ in type IIB string theory compactified on orientifolds of weakly-curved Calabi-Yau threefold hypersurfaces, the so-called  Kreuzer-Skarke axiverse \cite{Demirtas:2018akl}.   

The number of axions, $N$, in an effective theory is a very important parameter: both for the obvious reason that additional fields can have physical consequences, but also because axion couplings have been shown to display strong trends as functions of $N$ \cite{Demirtas:2018akl}.
Specifically, in the Kreuzer-Skarke axiverse,  
the axion decay constants $f$ become smaller as $N$ increases, while the couplings of some axions to the visible sector increase \cite{Demirtas:2018akl,superradiance,glimmers}.
One 
can therefore anticipate using experimental limits on axions to exclude certain classes of effective theories derived from string theory, \emph{for sufficiently large N} \cite{Halverson:2019cmy, superradiance,glimmers}.    

Because the abundance of axions in Calabi-Yau compactifications is a consequence of the rich topology of the internal space, rather than of dynamics that is specific to one weakly-coupled description, 
one might expect that the axiverses arising at different corners of the duality web should all be roughly comparable.
This expectation is incorrect: compactifications of F-theory \cite{Vafa:1996xn} to four dimensions on Calabi-Yau fourfolds host \emph{many} more axions than can occur in compactifications of  
superstring theory on Calabi-Yau threefolds.  
Specifically, compactifications of type IIB string theory on known Calabi-Yau threefolds yield at most $491$ four-form axions, 
whereas in compactifications of F-theory one can have as many as $181{,}820$ axions \cite{Yinan}.
 
Since $N$ can be extremely large in F-theory compactifications on Calabi-Yau fourfolds, it is natural to ask whether the resulting ensemble of axion theories, which we will refer to as the \emph{F-theory axiverse}, manifests any distinctive signatures. 
Previous studies of reheating \cite{Halverson:2019kna} and of the $N$-scaling of axion-photon couplings  \cite{Halverson:2019cmy} in the F-theory axiverse revealed 
complex phenomenology, but computational limitations  
restricted those studies to $N\leq 250$.

\medskip
The purpose of the present paper is to construct and analyze general incarnations of the F-theory axiverse, 
with the number of axions limited only by the existence of suitable fourfold topologies. 
The theories we study here are --- by a wide margin --- the largest axiverses known to admit ultraviolet completions. 
In this paper we make very significant computational advances 
that for the first time allow for the study of 
thousands of axions in F-theory compactifications.
In the same spirit as the analysis of type IIB compactifications on Calabi-Yau threefolds in \cite{Demirtas:2018akl}, 
the main task
is to assemble and carry out an algorithmic procedure to construct ensembles of axion theories: specifically, to open the door to theories with $250 < N \le 181{,}200$ axions. 
As such, this work largely involves applied computational geometry, and we leave detailed modeling of particle physics and cosmology as important goals for the future. 
In this way 
we prepare the ground for future studies of the rich physics of the F-theory axiverse.

\begin{figure}
    \centering 
    \includegraphics[width=0.95\linewidth]{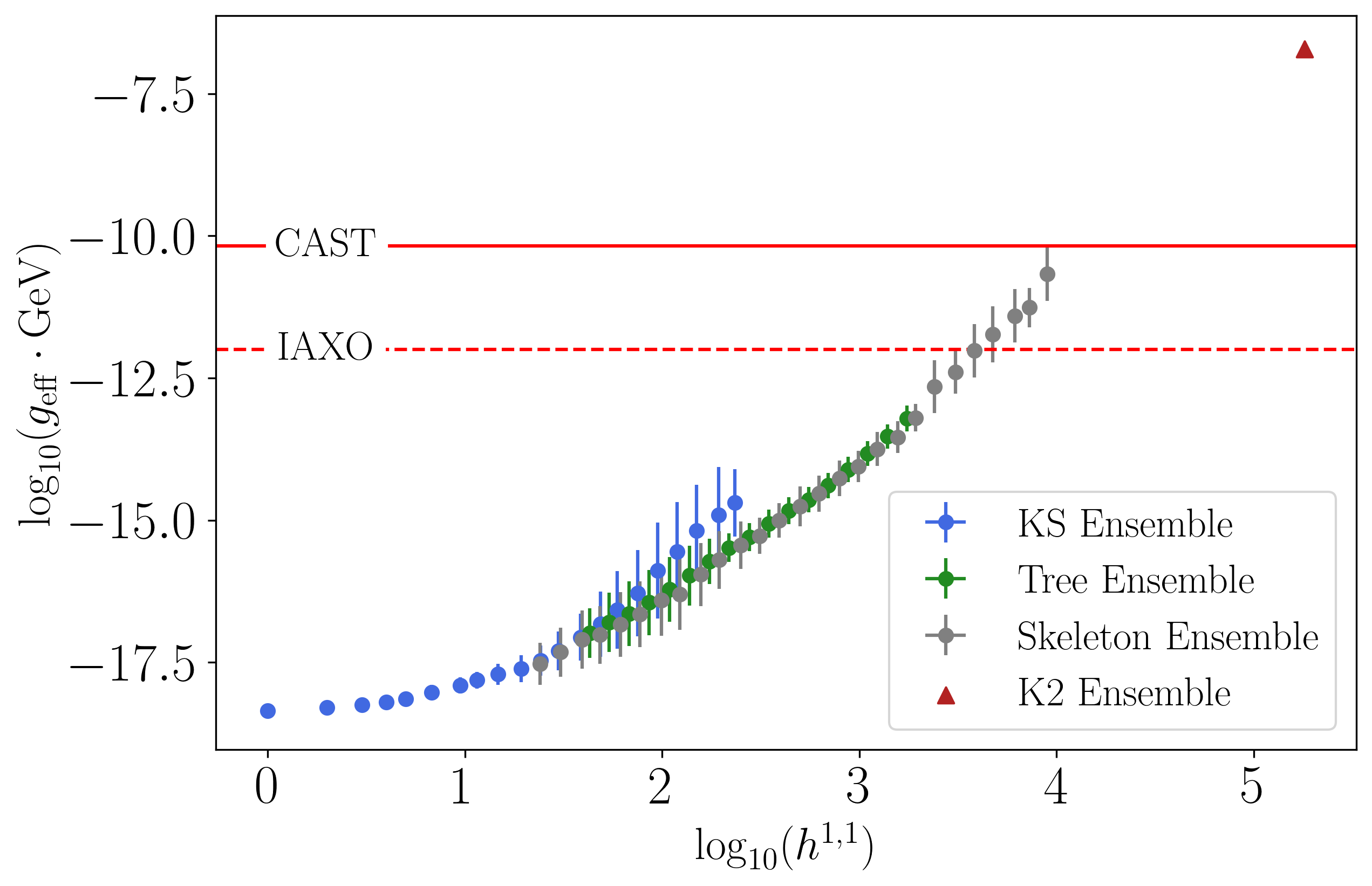}

\caption{The effective axion-photon coupling $g_{\text{eff}}$ as a function of 
the number of axions, $h^{1,1}$. 
The points show the mean value, and the error bars show the standard deviation in
each bin.
Models above the CAST line are excluded, and models above the projected IAXO line could be tested or excluded. 
We have imposed
$\alpha^{-1}_\text{EM, UV} \geq 19.5$, and applied a mass cut 
$m \le 10^{-2} \text{ eV}$ for comparison to helioscope experiments: see \S\ref{sec:methods}. The K2 point is at $h^{1,1}=181{,}200$, and the largest $h^{1,1}$ shown here for the Skeleton ensemble is $h^{1,1}=8{,}955$.}
    \label{fig:geff_h11}
\end{figure}

Specifically, we construct three different ensembles of four-dimensional theories from explicit fourfolds, following \cite{trees,skeleton,Yinan}, and --- making certain assumptions detailed below\footnote{In particular, we do \emph{not} explicitly stabilize moduli, but instead sample from a region in the moduli space where the $\alpha'$ expansion is plausibly controlled,
and where perturbative corrections to the K\"ahler potential could be large enough to make the saxions much heavier than the axions: see \S\ref{sec:methods}.} --- 
we compute the masses and decay constants of the resulting axions.  We also compute the axion couplings to gauge groups hosted on
geometrically non-Higgsable clusters (NHCs) \cite{Morrison:2012np,Morrison:2012js,Grassi:2014zxa,Morrison:2014lca,Halverson:2015jua, Halverson:2016vwx}.
Taking the NHCs as proxies for electromagnetism, our results provide a model of axion-photon couplings in F-theory. 
In this setting, we establish a rather precise correlation between the axion-photon couplings and $N=h^{1,1}$, and we find that current helioscope and X-ray limits already  exclude much of the model space in weakly-curved F-theory compactifications at large $h^{1,1}$.\\

\FloatBarrier

The organization of this paper is as follows.
In \S\ref{sec:ensembles} we introduce the compactifications that we will study. 
In \S\ref{sec:methods} we describe our assumptions and our methodology for computing axion couplings.
Our main results appear in \S\ref{sec:results}: we construct $64{,}485$ four-dimensional effective theories from F-theory compactified on Calabi-Yau fourfolds,\footnote{In addition, we study $33{,}506$ theories obtained from type IIB string theory compactified on Calabi-Yau threefolds.}  
and we report the statistics of the masses, decay constants,
and couplings to gauge fields
of the corresponding axions.
We conclude in \S\ref{sec:conclusions}.  Many technical details appear in the Appendices.

\section{Ensembles of F-theory Compactifications}\label{sec:ensembles}

In this section we define the three ensembles of F-theory compactifications constructed in this work.  Further details can be found in Appendix \ref{sec:appensembles}.

\subsection{Generalities of Weierstrass Models}

An F-theory compactification to four dimensions requires the specification of an elliptically fibered Calabi-Yau fourfold $X$. 
We take the base of the fibration to be a K\" ahler threefold $B$ and consider $X$ as a Weierstrass model
\begin{equation}
    y^2 = x^3 + f\, x + g\,,
\end{equation}
where $f$ and $g$ are polynomial in the coordinates of $B$: more precisely, $f\in \Gamma(\mathcal{O}(-4K_B))$ and $g\in \Gamma(\mathcal{O}(-6K_B))$. The elliptic fibers become singular along the discriminant locus $\Delta=0$, with
$\Delta = 4f^3+27g^2$. Loops around the discriminant locus yield Picard-Lefschetz monodromy on the middle homology of the elliptic fiber and an associated action on its complex structure, which is identified with an $SL(2,\mathbb{Z})$ action on the type IIB axiodilaton $\tau$, signalling the presence of seven-brane sources on $\Delta=0$. The na\"ive seven-brane gauge group is read off from the codimension one fibers --- as classified by Kodaira --- and may be further reduced via monodromy effects at codimension two.

Threefolds $B$ from the ensembles we study generally exhibit so-called \emph{geometrically non-Higgsable clusters} (NHCs). These occur when the most general form of $f$ and $g$ has overall factors along some base coordinate, e.g. $z$, as 
\begin{equation}
    f = z^a \, F \qquad g = z^b \, G \qquad \Delta = z^{c} \, \tilde \Delta\,, \qquad \qquad a,b \in \mathbb{N}\,,
    \label{eqn:NHC_def}
\end{equation} 
where $F$, $G$, and $\tilde{\Delta}$ are nonvanishing at $z=0$, and $c \ge  \text{min}(3a,2b)$.
In such a case there is  a seven-brane along $z=0$ for generic complex structure of $X$, and provided that $(a,b)\neq (1,1)$ the seven-brane carries a geometric gauge group. In situations where the gauge group on a seven-brane is obtained via tuning in moduli, there is a complex structure deformation that geometrically Higgses it. Since in  \eqref{eqn:NHC_def} there is no such deformation, we say that the seven-brane on $z=0$ is geometrically non-Higgsable, though 
the gauge group
could be broken by fluxes. The fiber and gauge data associated to NHCs are presented in Table \ref{tab:gauge_groups}.

The ensembles that we study are constructed by performing an appropriate tuning in the complex structure of $X$ and then doing a blowup of the base $B$ to $B'$, yielding a blowup of $X$ to $X'$, where $X'$ is elliptically fibered over $B'$. For this to be possible by traversing a finite distance in moduli space, the singularities along divisors $D$ (in $B$), curves $C$, and points $p$ must be sufficiently mild. Specifically,  
only if\footnote{The definition of the condition $\text{mult}_A(f, g) < (x, y)$ is that one or both of $\text{mult}_A(f) < x$ and $\text{mult}_A(g) < y$ holds, with $\text{mult}$ denoting the multiplicity on $A$.}
\begin{equation}
    \text{mult}_D(f,g) < (4,6)\,, \qquad \text{mult}_C(f,g) < (8,12)\,, \qquad \text{mult}_p(f,g) < (12,18)\,,
    \label{eqn:finite_distance}
\end{equation}
then one remains at finite distance in moduli space; see \cite{trees} for a discussion and the original mathematics references. In six-dimensional F-theory compactifications the condition we wrote for curves applies to points (both are codimension two in the base of their respective compactifications). There, points with multiplicity of vanishing between $(4,6)$ and $(8,12)$ play a crucial role in six-dimensional $\mathcal{N}=(1,0)$ SCFTs, where the curve in the base wrapped by the vanishing exceptional divisor of the resolution yields tensionless strings; see, e.g., \cite{Heckman:2015bfa}.
 
\subsection{F-theory Ensembles}

In this work we construct and study three different ensembles of fourfolds:
the Tree ensemble, following
\cite{trees}; the Skeleton ensemble, following \cite{skeleton}, and what we will call the K2 ensemble, following \cite{Yinan}.  We now provide brief definitions of each ensemble, referring the reader to the original papers \cite{trees, skeleton, Yinan} for more details.

\subsubsection{Tree Ensemble}\label{sec:tree}

The Tree ensemble introduced in \cite{trees} begins with a weak-Fano toric threefold $B_0$ obtained from a fine regular star triangulation of one of the 4,319 3d reflexive polytopes.  Elements of the ensemble are generated from this starting point in special sequences of blowups that are called ``trees",
since each individual blowup produces a new ray further from the origin than any in the cone being blown up in the original reflexive polytope. Each ray (or ``leaf") in the tree has $v= av_a + b v_b + c v_c\in \mathbb{Z}^3$ where $(v_a, v_b, v_c)$ are rays in the original polytope, and $h_v := a + b + c$ is 
called
the height of the leaf. The height of a tree is the height of its highest leaf. In \cite{trees} it was shown that the Hayakawa-Wang criterion \eqref{eqn:finite_distance} is satisfied provided that all trees have height $\leq 6$. This technical condition yields a systematic way to build a large ensemble of geometries: 1) classify all edge trees (associated to blowups of toric curves) and face trees (associated to blowups of toric points) with height less than or equal to six, 2) perform the sequences of blowups associated to uniformly sampled face trees, then 3) perform the sequences of blowups associated to uniformly sampled edge trees. This  yields a smooth K\"ahler threefold $B$ that serves as the base for a Weierstrass model. 

The possible bases $B$ in the Tree ensemble are dominated by those obtained from 
two 3d reflexive polytopes, which were denoted $\Delta^{\circ}_1$ and $\Delta^{\circ}_2$ in \cite{trees}.  In this work, we assemble bases built on $\Delta^{\circ}_1$ and $\Delta^{\circ}_2$, drawing half from each.

\subsubsection{Skeleton Ensemble}\label{sec:skeleton}

Another approach arises in the so-called Skeleton ensemble \cite{skeleton}, which performs sequences of random blowups\footnote{Alternatively, one can perform sequences of blowups and blow-downs \cite{skeleton}, but here we will perform a one-way process with only blowups.} beginning from a starting base chosen to be  $\mathbb{P}^3$. At any given step in this Monte Carlo process, one constructs the set of all possible toric blowups that satisfy a certain polytope condition \eqref{polytopecondition} that enforces equation \eqref{eqn:finite_distance}: 
see Appendix \ref{sec:appensembles} 
for details.
One then chooses one such blowup at random, from the uniform distribution on the allowed set.

\subsubsection{K2 Ensemble}\label{sec:K2}
Finally, 
in this work 
we introduce a new ensemble, following a construction by Wang in the approach to the F-theory geometry with the largest  
$h^{1,1}(B)$ \cite{Yinan}. That geometry $B_3$ is obtained by a sequence of toric blowups from a smooth threefold $B_\text{seed}$ to obtain a maximal-$h^{1,1}$ \emph{toric} base $B_\text{toric}$, followed by a further sequence of non-toric blowups to obtain $B_3$: see Figure 1 of \cite{Yinan}.
These have 
\begin{equation}
    h^{1,1}(B_\text{seed}) = 2{,}560\,, \qquad h^{1,1}(B_\text{toric}) = 181{,}200\,, \qquad h^{1,1}(B_3) = 181{,}819\,.
\end{equation}
Keeping with our natural theme: if with respect to $h^{1,1}$, $B_3$ is Everest, then $B_\text{toric}$ is K2.
In performing the sequence of blowups to obtain $B_\text{toric}$, one arrives at an ensemble of intermediate bases $B$ that we call the K2 ensemble. 

Specifically,
if face trees are planted on all 5,016 toric (8,12) points of $B_\text{seed}$ and edge trees are then planted on all 7,576 toric (4,6) curves of $B_\text{seed}$, one obtains a fan corresponding to a compact, smooth toric threefold $B_\text{tree}$. Since the rays of $B_\text{tree}$ are a subset of the rays of $B_3,$ and since $B_3$ satisfies the hypotheses of Proposition \ref{prop:poly}, then $B_\text{tree}$ satisfies the hypotheses of Proposition \ref{prop:poly}, and so \eqref{eqn:finite_distance} holds on all toric subvarieties of $B_\text{tree}$. We define the K2 ensemble to be $B_3$ with the triangulation described in \S2 of \cite{Yinan}, along with the collection of toric threefolds obtained by adding adding face trees to all 5,016 toric (8,12) points of $B_\text{seed}$ then planting edge trees on all 7,576 toric (4,6) curves of $B_\text{seed}$. This ensemble consists of $5.11 \times 10^{52{,}730}$ geometries, from which one can draw uniformly at random.\footnote{Note, however, that this only yields $1.78 \times 10^{32{,}924}$ geometries at $h^{1,1} = 181,200$ instead of the $7.5 \times 10^{45{,}766}$ obtained in \cite{Yinan}, because with the exception of the triangulation in \S2 of \cite{Yinan}, we restrict to adding face trees before edge trees.}\footnote{This count assumes there are no nontrivial $\text{GL}_{N}(\mathbb{Z})$ fan automorphisms.}

\section{Computation of Axion Couplings}\label{sec:methods}

To build an ensemble of four-dimensional effective theories, we begin with the  Calabi-Yau fourfolds constructed in \S\ref{sec:ensembles}, each of which
has a moduli space of Ricci-flat metrics.
The next step is to specify one or more points in the moduli space of each compactification, and then compute the four-dimensional theory at each such point.  

In \S\ref{sec:precis} we lay out the logic for this computation, explaining how we choose points, and what physical approximations we make.
Then, in \S\ref{sec:comp}, we explain the computational procedure in detail.

\subsection{Assumptions}\label{sec:precis}
\vspace{5pt}

We now briefly summarize the assumptions that we make in 
order to assemble a set of axion effective theories. 

Computing the dynamics of moduli stabilization in general F-theory compactifications is currently out of reach, and so we \emph{select} points.  Specifically, we start by taking the K\"ahler moduli to be at the tip of the stretched K\"ahler cone, as defined carefully below.
We make this choice in order to control corrections in the $\alpha'$ expansion, and also to model the expected effects of perturbative moduli stabilization.
We then perform a series of adjustments of four-cycle volumes to produce a proxy for electromagnetism: we ensure that at least one NHC has a gauge coupling in a range specified below.  Each resulting point in K\"ahler moduli space defines a model.

For each such model, we approximate the four-dimensional effective action as follows.  We omit perturbative and nonperturbative corrections to the K\"ahler potential, and we omit St\"uckelberg couplings to gauge fields.  For each NHC, we include the gaugino condensate superpotential term, and for each prime toric divisor that does not host an NHC, we include a Euclidean D3-brane superpotential term with unit Pfaffian.  We truncate the nonperturbative superpotential to a basis set of the $h^{1,1}$ leading terms.  Finally, we take the vev of the Gukov-Vafa-Witten flux superpotential to be $W_0=1$.
In sum, in our F-theory compactifications a model is specified by a choice of a Calabi-Yau fourfold, of an NHC standing in for electromagnetism, and of a point in K\"ahler moduli space
where the volume of the cycle hosting the NHC has a specified value.\footnote{In the weakly-coupled type IIB compactifications that we present for comparison purposes, a model is specified by a choice of Calabi-Yau threefold topology, of a prime toric divisor on which D7-branes could host electromagnetism, and of a value for the volume of the corresponding cycle.}

\bigskip
For the reader who wishes to understand the logic behind the simplifying assumptions that we have just outlined, we will now lay out our reasoning more completely.

We suppose that the complex structure moduli and axiodilaton are stabilized at a high scale by fluxes.  Stabilization of the K\"ahler moduli by nonperturbative superpotential terms would induce parametrically comparable masses for the saxions and the axions, and if this scale were high enough to prevent cosmological problems from the saxions, the axions would likewise be heavy enough to decouple from physical processes occurring after nucleosynthesis.  For our purpose of studying 
the phenomenology of light axions, this is an uninteresting outcome, and we therefore consider parameter regimes where other moduli stabilization mechanisms are operative.

Specifically, we have in mind stabilization of the real-part K\"ahler moduli (i.e., saxions) by \emph{perturbative} corrections to the K\"ahler 
potential.   Such corrections do not depend on the axions, whose shift symmetry is broken only nonperturbatively, and hence in the resulting vacua the saxions can become massive while leaving the axions light.  Explicitly constructing vacua along these lines is a challenge for the future. 

In the spirit of the Dine-Seiberg problem \cite{Dine:1985he}, we expect that stabilization by corrections that are perturbative in the $\alpha'$ expansion will lead to vacua that are neither parametrically weakly curved, nor 
parametrically strongly curved, but in which instead some curves have volumes of order unity.   
Such vacua would lie close to the boundary of the region where the $\alpha'$ expansion is well-controlled.\footnote{For an alternative mechanism that might lead to stabilization in a similar parameter regime, see \cite{Casas:2025bvk}.}

With this in mind, we make a very significant assumption, whose consequences propagate through all our findings: we suppose that the K\"ahler moduli are stabilized 
\emph{inside the stretched K\"ahler cone.}
Recalling that the K\"ahler cone $\mathscr{K}$ is parameterized by K\"ahler parameters (curve volumes) $t_i$, as
\begin{equation}
\text{K\"ahler cone }\mathscr{K}:~~\Bigl\{\vec{t}\,\Bigr|~\vec{t}\cdot\vec{C}\ge 0~\forall \vec{C} \in \mathcal{M}\Bigr\}\,,
\end{equation} with $\mathcal{M}$ the Mori cone, for a real number $r>0$
we define  the $r$-stretched K\"ahler cone as
\begin{equation}
r\text{-stretched K\"ahler cone } \mathscr{K}_{r}:~~\Bigl\{\vec{t}\,\Bigr|~\vec{t}\cdot\vec{C}\ge r~\forall \vec{C} \in \mathcal{M}\Bigr\}\,.
\end{equation}
In this work we consider compactifications whose K\"ahler moduli lie in the $1$-stretched K\"ahler cone, $\mathscr{K}_{1}$,
and that also obey a further  quadratic constraint given in \eqref{eq:dcskc_def} related to divisor volumes and gauge instanton contributions to the superpotential.  

More specifically, we study points at or near the 
the tip of $\mathscr{K}_{1}$, defined as the locus in  $\mathscr{K}_{1}$ with smallest distance from the origin.\footnote{Depending on the notion of distance used in the K\"ahler moduli space, one defines different tips.
For the Tree ensemble defined in \S\ref{sec:tree}, the Kreuzer-Skarke ensemble with $h^{1,1} \leq 174$, and the Skeleton ensemble (\S\ref{sec:skeleton}) with $h^{1,1} \leq 2{,}207$, we use the $\ell^2$ tip resulting from the $\ell^2$ norm. Otherwise, due to the computational expense of computing the $\ell^2$ tip, for the Kreuzer-Skarke ensemble with $h^{1,1} \geq $175 and the Skeleton ensemble with $h^{1,1} \geq 2{,}208$, we use the $\ell^1$ tip. For the K2 ensemble at $h^{1,1} = 181{,}200$, even finding the $\ell^1$ tip is too expensive, so we find a point in the stretched K\"ahler cone and then dilate it until it reaches the dual Coxeter stretched K\"ahler cone.}  
Within $\mathscr{K}_{1}$, the compactification volume $\mathcal{V}$ 
typically takes its smallest value near\footnote{The $\ell^1$ distance from the origin in K\"ahler moduli space is of course not the same function of K\"ahler moduli as the volume $\mathcal{V}$, but the latter is cubic and much more costly to minimize than the $\ell^1$ distance.} the tip. Because axion-photon couplings generally increase as $\mathcal{V}$ increases, the tip is thus an appropriate and easily-computed reference location for estimating a lower bound on axion-photon couplings.  To reiterate this key point: with high probability, a point in K\"ahler moduli space where the $\alpha'$ expansion is \emph{better} controlled than it is at the tip will produce a model that is \emph{more} constrained by axion-photon limits than the tip model is.  Thus, we can obtain conservative bounds by working at the tip.

Let us stress that we have not provided a dynamical explanation for stabilization near the tip.  Instead, we have argued that perturbative corrections in the $\alpha'$ expansion plausibly lead to stabilization near a wall of $\mathscr{K}_{r}$ for some $r \lesssim \mathcal{O}(1)$, and for concreteness we take $r=1$ henceforth. Understanding the dynamics of perturbative moduli stabilization is an important problem for future work.

Having defined a location in moduli space, we turn to computing the effective action.
The axions resulting from reducing the Ramond-Ramond four-forms on 4-cycles can acquire mass from a variety of sources:
\begin{enumerate}    
    \item nonperturbative contributions to the superpotential, from 
    Euclidean D3-branes and from gaugino condensation in super-Yang-Mills sectors.
    \item nonperturbative contributions to the K\"ahler potential, likewise from  Euclidean D3-branes and gaugino condensation.
    \item St\"uckelberg couplings to gauge fields.
\end{enumerate} 
St\"uckelberg masses can be induced by certain fluxes (see e.g.~\cite{Grimm:2011dj,Grimm:2011tb}), but a proper study of such effects is beyond the scope of this work.
We will 
neglect St\"uckelberg couplings in this paper, aside from offering the following estimate. 
Letting $G$ be the geometric gauge group on the NHCs, we expect that 
the number $N$ of axions that do not participate in St\"uckelberg couplings is roughly
\begin{equation}\label{eq:stuck}
    N \ge h^{(1,1)}(B) - \text{rk}(G) \simeq \frac{h^{1,1}(B)}{2},
\end{equation}
where in the estimate we have used a rough empirical result from typical geometries in our ensembles at large $h^{1,1}$.

Turning to nonperturbative corrections, we make a simplifying assumption: we include a superpotential term for
each prime toric divisor, but not for any other divisors, and we omit nonperturbative contributions to the K\"ahler potential.
In Appendix \ref{app:couplings} we explain the logic behind this \emph{prime toric model}, and we comment on the expected accuracy of the associated approximation.

One particular superpotential term has a critical effect in our analysis of axion-photon couplings:  this is a 
term from the stack of seven-branes that we use as a proxy for the visible sector.\footnote{We assume that flux breaks this particular gauge group to a product that contains a $U(1)$, which we then use as a stand-in for electromagnetism.}
Specifically, writing $D_{\text{SM}}$ for the associated divisor,
and $T_{\text{SM}}$ for its complexified volume,
we consider a Euclidean D3-brane superpotential term
\begin{equation}
    W \supset e^{-2\pi T_{\text{SM}}}\,,
\end{equation}
which we refer to as a \emph{QED instanton}.   
As explained in \cite{glimmers}, such a term introduces a new mass scale $m_{QED}$, called the \emph{light threshold} in \cite{glimmers}, and axions with masses $m \ll m_{QED}$ suffer hierarchically suppressed couplings to the photon.\footnote{
The corresponding effect for GUTs was laid out in the earlier paper \cite{Agrawal:2022lsp}.  The light threshold effect as discussed in \cite{glimmers} is a stringy generalization, in which D-brane instantons contribute terms that are not necessarily found in the low-energy gauge theory.}

At various points in our analysis we will consider the effect of such QED instantons. Multiple lines of argumentation suggest that such BPS instantons associated to $U(1)$ gauge groups exist in string theory, even if they sometimes have additional fermion zero modes that forbid a superpotential correction. For instance, if the $U(1)$ arises from a seven-brane on a divisor\footnote{In more realistic models QED does not wrap such a simple divisor, but arises from combined effects related to electroweak symmetry breaking and associated mixing.} $D$, a Euclidean D3-brane on $D$ yields the requisite instanton. Additionally, we may understand a $U(1)$ instanton from the point of view of symmetry breaking. Though $U(1)$ does not support a topologically non-trivial smooth gauge field configuration, $SU(2)$ does. If that $SU(2)$ is broken to $U(1)$ by an adjoint then the expectation value of the Higgs field generates a potential for the size modulus of the $SU(2)$ instanton via a 't Hooft term \cite{PhysRevD.14.3432} see, e.g., \cite{Halverson_2016} for a discussion. That potential is minimized not by a smooth gauge field configuration, but by a small instanton, which is indistinguishable in string theory from a brane within a brane \cite{douglas1995branesbranes, Witten_1996}. That is, there is no smooth gauge field configuration realizing the small $SU(2)$ instanton of the theory after breaking to $U(1)$, but we know that string theory can accommodate this configuration. Finally, even if the $U(1)$ is not realized by a supersymmetric Higgsing of $SU(2)$ (or a higher group), one might make a similar argument in a brane/anti-brane system, via tachyon condensation.  For these reasons we find it plausible that the mentioned QED instanton effect is present in string theory, and it may correct the superpotential if it has the right fermion zero modes.

We have argued above that the presence of QED instantons is plausible, and hence we include their effects in all our computations of the effective axion-photon coupling $g_{\text{eff}}$, as defined in \eqref{eq:geff}.
In particular, the results for $g_{\text{eff}}$ shown in Figure \ref{fig:geff_h11} are obtained with the assumption that QED instantons exist and contribute to the axion potential.
However, we stress that the effect of such
QED instantons is to \emph{reduce} $g_{\text{eff}}$  
\cite{glimmers}.
Thus, if QED instantons were omitted, we would find stronger constraints from helioscopes such as CAST and IAXO.

\subsection{Computational Procedure}\label{sec:comp}

\subsubsection{Axion Couplings from F-theory}

Let $B_3$ be a smooth compact three-dimensional K\"ahler manifold with $h^{1,1}(B_3) = N$. In the case of the Tree, Skeleton, and K2 ensembles, $B_3$ is a toric threefold, whereas in the Kreuzer-Skarke ensemble, $B_3$ is a Calabi-Yau hypersurface in a four-dimensional toric variety.  We choose a convenient basis of divisor classes $[D_1], \ldots, [D_N]$  for $H_4(B_3, \mathbb{Z})$, and denote
the holomorphic representatives by $D_a$, $a=1,\ldots,N$.
For a given K\"ahler form $J$, we have the volumes
\begin{equation}
    \tau^a := \frac{1}{2}\int_{[D_a]}J\wedge J\,.
\end{equation} 
The complexified volumes
\begin{equation}
    T^a:= \tau^a+i\,\theta^a\,,
\end{equation}
with
\begin{equation}
    \theta^a := \int_{[D_a]} C_4\,,
\end{equation}
serve as coordinates on K\"ahler moduli space.

We will now write down a formal general expression for the superpotential for the $T_i$, and then explain 
the assumptions we make about nonperturbative contributions in order to evaluate the formal expression.
Let $D_A$ be a divisor whose class is $[D_A] = \sum_{a} Q_{Aa}[D_a]$, with $Q_{Aa} \in \mathbb{Q}$ and with $A \in \mathbb{N}$.
If $D_A$ hosts a non-Higgsable gauge group,
we write $C_2(D_A)$ to denote the dual Coxeter number of the gauge group residing on $D_A$, and for all other $D_A$ we define $C_2(D_A)=1$.
We can thus write the
full superpotential as
\begin{equation}
\label{eq:wfull}
W  = W_0 + \sum_{D_A} \mathcal{A}_A\,  e^{-2 \pi Q_{Ab}T^b/C_2(D_A)}\,,
\end{equation}
where $W_0$ is the Gukov-Vafa-Witten flux superpotential.
The sum in \eqref{eq:wfull} in principle runs over the infinite set of all effective divisors $D_A$ in $B_3$, not just the $N$ basis divisors $D_a$, but for some effective divisors the number of fermion zero modes may exceed two, precluding a superpotential contribution.
This case can be handled by taking the Pfaffian prefactor\footnote{See \cite{Alexandrov:2022mmy} for recent progress in characterizing the overall normalization of Euclidean D-brane Pfaffians, and \cite{Cvetic:2011gp, Cvetic:2012ts} for the complex structure moduli dependence.}
$\mathcal{A}_A$ to vanish for such non-contributing $D_A$.
Nonzero contributions from $D_A$ that host non-Higgsable gauge groups result from gaugino condensation, while nonzero contributions from other $D_A$ result from Euclidean D3-branes wrapping $D_A$.

To replace the sum in \eqref{eq:wfull} with a tractable finite sum, we note that for all but finitely many $A$, the volume $Q_{Aa}\tau^a$ will be so large as to make the corresponding superpotential term negligible.  A convenient simplification that suffices for computing most properties of the axion effective theory, and that we will employ throughout, is to truncate the sum over instantons to the $N$ largest contributions.

Specifically, we assume a \emph{prime toric model}, in which every prime toric divisor hosts either an NHC or a Euclidean D3-brane superpotential term.
There are $h^{1,1}+3$ prime toric divisors in a toric threefold, and $h^{1,1}+4$ prime toric divisors in a Calabi-Yau hypersurface in a toric fourfold.  As we explain in Appendix \ref{app:couplings},
the $N=h^{1,1}$ linearly independent prime toric divisors with the smallest instanton actions typically give a good approximation to the leading contributions overall.
We therefore choose our basis elements $[D_1],\ldots,[D_N]$ to be exactly these $N$ dominant prime toric divisors.\footnote{In general, $[D_1],\ldots,[D_N]$ may furnish a $\mathbb{Q}$-basis for $H_4(B_3, \mathbb{Z})$ rather than a $\mathbb{Z}$-basis, cf.~Appendix \ref{app:couplings}.}  
Moreover, the NHCs of interest arise on a subset of the $D_a$.  We use
the index $\alpha$ to denote the subset of $\{1,\ldots,N\}$ for which $D_\alpha$ hosts a non-Higgsable gauge group, so  
$\{\alpha\} \subsetneq \{a\}$. Evidence supporting this model in the Tree ensemble is provided in \cite{Halverson:2019vmd}.

In sum, restricting to the leading prime toric divisors and taking these same divisors as basis elements, the superpotential is
\begin{align}
W = W_0 + \sum_{a=1}^N \mathcal{A}_a\,  e^{-2 \pi T^a/C_2(D_a)}\,.
\end{align} 
With  appropriate shifts of the $\theta^a$, we can absorb any complex phases in the Pfaffians $\mathcal{A}_a$.

In F-theory, we have \cite{Grimm:2010ks}
\begin{align}K & = -2 \log \mathcal{V} + K_\text{other}\,,\label{eq:kahler_potential}
\end{align} 
with
$\mathcal{V}$ denoting the volume of $B_3$.
As a rough model for $K_\text{other}$,
we take 
\begin{equation}\label{eq:kother}
K_\text{other} = \log \bigg(\frac{g_s^4}{128}\bigg)\,,
\end{equation} 
as in \cite{fuzzy}.
The K\"ahler potential receives corrections to all orders in $g_s,$ and 
because there are strongly-coupled regions near most divisors hosting non-Higgsable gauge groups (see e.g.~\cite{Halverson:2017vde}), we stress that \eqref{eq:kahler_potential} in all likelihood is not a good model for the exact K\"ahler potential in F-theory.

The part of the supergravity Lagrangian relevant to axion physics takes the form 
\begin{equation}\label{eq:sugra}
\mathcal{L} \supset -\frac{1}{2} \mathcal{K}_{ab} \partial_\mu \theta^a \partial^\mu \theta^b - \sum_{a=1}^N \Lambda_a^4 \cos\bigl(2 \pi \theta^a\bigr) - \sum_\alpha  
\Bigl(\tau^\alpha G_{\alpha\mu\nu}G^{\;\mu\nu}_\alpha + \theta^\alpha G_{\alpha \mu\nu}\tilde{G}_\alpha^{\;\mu\nu}\Bigr),
\end{equation} 
in units where the reduced Planck mass is set to one. 
In \eqref{eq:sugra},
$\mathcal{K}_{ab}$ is the K\"ahler metric, and 
\begin{equation}\Lambda_a^4 = 8 \pi W_0 \times \frac{\mathcal{A}_a\tau^a}{\mathcal{V}^2}e^{-2 \pi \tau^a/C_2(D_a)} e^{K_\text{other}},\end{equation} are the instanton scales.

In the remainder of this work, we will set the Pfaffians $\mathcal{A}_a$ and the Gukov-Vafa-Witten superpotential $W_0$ to unity, as explained in \S\ref{sec:precis}.  

\subsubsection{Algorithm for Computing $(m,f)$}
 
We now redefine the axion fields to achieve the following:
\begin{enumerate} 
    \item Canonicalize the kinetic term\,,
    \item Diagonalize the mass term\,.
\end{enumerate}
We first perform 
a Cholesky decomposition, 
\begin{equation}({\mathcal{K}}^{-1})^{ab} = ( {L} {L}^T)^{ab},
\end{equation}  
where $ {L}^a_{\; b}$ is a lower-diagonal $N \times N$ matrix with positive diagonal 
entries.
Then we can define\footnote{We work with $M_{\text{pl}}=1$ for most of this paper, but include factors of $M_{\text{pl}}$ in a few places for emphasis.} \begin{equation} {\theta^a} = \frac{1}{M_\text{pl}}   {L}^a_{\; b} \phi^b,\end{equation} so that the $\phi^a$ 
have canonical kinetic terms, but a non-diagonal mass matrix $M_{ab}$.
We then diagonalize the  mass matrix by an orthogonal transformation $H$,
\begin{equation}
    \text{diag}(m_a^2) = H^T M H\,,
\end{equation}
and write
\begin{equation}
\phi^c := H^c_{\; d} \varphi^d,
\end{equation}
so that 
\begin{equation}
    \phi^a M_{ab} \phi^b = m_a^2 (\varphi^a)^2\,.
\end{equation}
In sum, the mass-and-kinetic eigenstate fields  $\varphi^a$ are written in terms of the original axion fields $\theta^c$ as  
\begin{equation}\varphi^a :=   (H^T)^a_{\; b}(L^{-1})^b_{\; c}   \theta^c\,.\end{equation}

After canonically normalizing the gauge fields, the Lagrangian 
takes the form  
\begin{align}
\mathcal{L}  \supset & -\frac{1}{2}  \bigg(\partial_\mu \varphi^a \partial^\mu \varphi^a + m_a^2 (\varphi^a)^2\bigg) + \frac{1}{4!}  \lambda_{abcd} \varphi^a \varphi^b \varphi^c \varphi^d + \mathcal{O}(\varphi^6) \nonumber\\ & - \frac{1}{4}\left(F_{\mu\nu}F^{\mu\nu} + c_{a} \varphi^a F_{\mu\nu} \tilde{F}^{\mu\nu}\right) - \frac{1}{4} \sum_{\alpha \neq \text{EM}} \bigg(G_{\alpha\mu\nu} G_\alpha^{\;\mu\nu} + c_{\alpha, a} \varphi^a G_{\alpha\mu\nu} \tilde{G}_{\alpha}^{\;\mu\nu}\bigg),
\end{align} 
where the quartic couplings are given by 
\begin{equation}\lambda_{abcd}  = (2 \pi)^4\sum_{m = 1}^N \Lambda^4_m (LH)_{ma} (LH)_{mb} (LH)_{mc}(LH)_{md}\,,
\end{equation} 
and the coupling of the $b$th mass eigenstate axion field to the $\alpha$th gauge group is given by
\begin{align}c_{\alpha,b} & = \frac{1}{M_\text{pl} \tau^\alpha} (LH)_{\alpha b} \qquad \text{(no sum)}\,,\label{eq:axion_gauge_group_coupling}\end{align} 
and we have introduced the notation $F_{\mu\nu}$ for the field strength of electromagnetism, and $G_{a\mu\nu}$ for the field strengths of other gauge groups indexed by $a$.   

In this setting it is not obvious how to define a decay constant $f_i$ for each axion. We do so by analogy with the single axion case, following \cite{superradiance}.  In the case of a single axion $a$ in a simple cosine potential, 
one has 
\begin{align}
\mathcal{L} & \supset  -\frac{1}{2} \bigl( \partial_\mu a \partial^\mu a  + m^2 a^2 \bigr) + \frac{1}{4!} \frac{m^2}{f^2}a^4 + \ldots \nonumber\\ & - \frac{1}{4}\left(F_{\mu\nu} F^{\mu\nu} + g_{a\gamma\gamma} a F_{\mu\nu} \tilde{F}^{\mu\nu}\right)\,,
\end{align} 
where $m$ is the mass of $a,$ $f$ is its decay constant, and $g_{a\gamma\gamma}$ is its axion-photon coupling. In the multi-axion setting, then, we will define the decay constant of the $a$th axion $\varphi^a$ to be \begin{align}\label{eq:quarticdef}
f_a := \sqrt{\frac{m_a^2}{\lambda_{aaaa}}}\,.\end{align}  
In the hierarchical limit, where $\Lambda^4_a \gg \Lambda^4_{b}$ for all integers $1 \leq a < b \leq N$, the definition \eqref{eq:quarticdef} coincides with the the result of \cite{glimmers}.

\subsubsection{Couplings to gauge sectors}

Equation \eqref{eq:axion_gauge_group_coupling} gives the coupling of the $b$th mass eigenstate axion field to the $a$th gauge group in the UV. We can rephrase this in terms of the UV gauge coupling $\alpha_\beta$ using \begin{align}\alpha_\beta & := \frac{1}{\tau^\beta}\,,\end{align} which allows us to rewrite equation \eqref{eq:axion_gauge_group_coupling} as \begin{align}c_{\beta, b} & = \frac{\alpha_\beta}{M_\text{pl}}(LH)_{\beta b}\,. \label{eq:em_coupling_for_later}\end{align} 
Further precision could be added by including the Weinberg angle to relate the gauge couplings of $SU(2)_L\times U(1)_Y$ to that of $U(1)_\text{EM}$; see \cite{Halverson:2019cmy} for a treatment in F-theory.

We wish to compare these couplings to experiments 
that probe axion-photon couplings, including CAST \cite{CAST:2017uph}, Chandra (see e.g.~\cite{Berg:2016ese,Reynes:2021bpe}), IAXO \cite{IAXO:2019mpb}, and STROBE-X \cite{ray2019strobexxraytimingspectroscopy}
In the absence of an explicitly constructed Standard Model, we assume that electromagnetism is realized as a subgroup of a non-Higgsable gauge group. To ensure that the gauge coupling can potentially flow to the right value in the IR, namely $\alpha_\text{EM} \simeq \frac{1}{137}$, we require the volume of the divisor hosting the non-Higgsable gauge group containing electromagnetism to satisfy $\tau^{\text{EM}} \leq 50$ \cite{Halverson:2019cmy,glimmers}.

The other complication is that experiments are only sensitive to the axion-photon couplings of axions with masses $m_i < m_\text{cutoff},$ where $m_\text{cutoff}$ varies by experiment. For helioscopes such as CAST and IAXO \cite{CAST:2017uph,IAXO:2019mpb}
one has  
$m_\text{cutoff} \approx 10^{-2} \text{ eV},$ while for the Chandra and STROBE-X X-ray experiments, one has $m_\text{cutoff} \approx 10^{-12} \text{ eV}$ \cite{Reynes:2021bpe, ray2019strobexxraytimingspectroscopy}.  
To compare with the constraints from these experiments, we add 
in quadrature the axion-photon couplings from all axions with masses below the cutoff, so for a given experiment and choice of non-Higgsable gauge group with 
charges $Q^\text{EM}_i$, the effective axion-photon coupling is given by \begin{align}
g_\text{eff} & = \sqrt{\sum_{ m_a < m_\text{cutoff}} c_{\text{EM}, a}^2} \\ & = \frac{1}{137} \frac{1}{M_\text{pl}} \sqrt{Q^\text{EM}_b Q^\text{EM}_c\sum_{ m_a < m_\text{cutoff}} (LH)_{ba} (LH)_{ca}}\,.\label{eq:geff}\end{align} 
When $m_\text{cutoff}\to \infty$, we obtain 
\begin{equation}
    g_\text{eff, no cutoff} = \frac{1}{137} \frac{1}{M_\text{pl}} \sqrt{(K^{-1})_{\text{EM, EM}}}\,,
    \label{eq:geff_nocut}
\end{equation}
which we find in practice is a good approximation to \eqref{eq:geff} when $L$ has one large entry.
For instance, we tested this approximation in the first $19$ bins of the Skeleton ensemble by finding the first example that has $L$ with one large entry, finding that the maximum absolute percent error of \eqref{eq:geff} relative to \eqref{eq:geff_nocut} is $(1.54\times 10^{-6})\%$. We also matched this error to the theoretically expected errors, in most examples up to an $\mathcal{O}(1)$ factor.

\subsubsection{Numerical Methods} 

The hardest part of the above computation is finding the special orthogonal matrix $H$ diagonalizing the mass matrix. 
The principal difficulty is that the ratio of the largest instanton action to the smallest instanton action is often $\mathcal{O}(10^4)$, which upon exponentiation leads to enormous hierarchies.
As a result, extremely high precision is required.

Carrying out the diagonalization at such precision using e.g.~\texttt{mpmath}, while possible in principle, becomes 
intractable after $h^{1,1} \sim 200$ from both memory and time considerations. 
For example, with $\mathcal{O}(10^3)$ axions, finding the eigenvalues and eigenvectors exactly would require keeping at least $\mathcal{O}(10^{10})$ doubles in computer memory, requiring very roughly $\mathcal{O}(100 \text{ GB})$ of RAM. 
Moreover, since eigenvalue calculations of an $N\times N$ matrix scale (roughly) as $N^3$, we estimate from our calculations with $200$ axions that the calculation for $1000$ axions would take at least a month. This is beyond the time limit  of our cluster. We emphasize that this calculation is for matrices with large hierarchies requiring very high precision calculations.
In sum,
brute-force computation of $H$ to 
sufficient numerical precision is out of reach for part of the regime of interest to us.

Fortunately, in cases where the instanton scales are hierarchical, i.e.~$\Lambda_i^4 \gg \Lambda_{i + 1}^4$ for all $i$, one can successively integrate out axion fields, as in  \cite{glimmers}, and this determines the components of $H$ to be \begin{equation}H^a_{\; b}= \delta^a_{\; b} + \mathcal{O}\bigl(\Lambda^4_{\min (a, b)} / \Lambda^4_{\text{max}(a,b)}\bigr)\,,\end{equation} where the subleading terms can be determined via standard perturbation theory techniques.  
However, in the K2, Skeleton, and Tree ensembles, the instanton scales are often not hierarchical. Instead they come in hierarchically separated groups of nearly degenerate scales, whose geometric origin is a pattern of degeneracy in divisor volumes. 
Since the cosmologically relevant axions are those that have rolled since the beginning of the universe, i.e. those axions with mass $m \gtrsim 3 H_0,$ where $H_0$ is the present-day Hubble parameter, we use \texttt{mpmath} to determine the components of $H^a_{\; b}$ corresponding to such axions, and use perturbative methods for the ultra-ultralight axions that remain. 

In the K2 ensemble, we computed a point in the stretched K\"ahler cone of $B_3,$ the toric threefold base constructed in \S2 of \cite{Yinan}. Then we dilated the point by a factor of three so that it was contained in the dual Coxeter stretched K\"ahler cone. We selected one of the seven divisors hosting a non-Higgsable gauge group with $|\text{vol}(D)-30| \le 1$
to host electromagnetism. We then used equation \eqref{eq:geff_nocut} to compute $g_{\text{eff}}$,\footnote{For our choice of EM divisor the Cholesky factor $L$ had $L_{\text{EM, EM}} = 6.46 \times 10^{13},$ and the next largest element of $L_{\text{EM, }a}$ in absolute value was $8.7 \times 10^{-11}$, justifying the use of \eqref{eq:geff_nocut}.}
since using \eqref{eq:geff} was computationally intractable, due to the large number of axions in the sum.


\subsection{Construction of the Ensembles}\label{sec:ensemblesummary}

We now specify the choices that we made in sampling geometries.
First we chose bins for $h^{1,1}$, spaced by roughly $0.1$ on a $\log_{10}$ scale.
With $[a, b]$ indicating that we took geometries with $h^{1,1}$ between $a$ and $b$ inclusive, we used following $h^{1,1}$ bins:\\
$\Bigl\{$[1, 1], [2, 2], [3, 3], [4, 4], [5, 5], [6, 7], [8, 10], [11, 12], [13, 16], [17, 21], [22, 26], [27, 34], [35, 43],    [44, 54], [55, 68], [69, 86], [87, 109], [110, 138], [139, 174], [175, 219], [220, 277], [278, 349],   [350, 439], [440, 553], [554, 697], [698, 878], [879, 1105], [1106, 1392], [1393, 1753], [1754, 2207]   [2208, 2779], [2780, 3499], [3500, 4405], [4406, 5546], [5547, 6982], [6983, 8790], [8791, 11066],  [11067, 13932], [13933, 17540]$\Bigr\}$.

To collect a reasonably fair sample of Calabi-Yau threefold hypersurfaces that are not trivially equivalent, we sampled non-two-face-equivalent fine, regular, star triangulations (NTFE FRSTs) using the code and methods developed in \cite{macfadden2023efficientalgorithmgeneratinghomotopy}. At $h^{1,1} = 1, 2$ in the Kreuzer-Skarke database, there are only 5 and 39 NTFE FRSTS of $N$-favorable polytopes, and so we took all of them. For the $h^{1,1}$ bins with $h^{1,1}$ between 3 and 349, we sampled 100 $N-$favorable polytopes uniformly at random from the set of all polytopes with the given $h^{1,1}$ and then took their Delaunay triangulations. For $h^{1,1}$ bins with $h^{1,1}$ between 350 and $491,$ since there were fewer than 100 polytopes per bin, we took all $k$ $N-$favorable polytopes in that $h^{1,1}$ bin and used the \verb|random_triangulations_fast()| method from \cite{macfadden2023efficientalgorithmgeneratinghomotopy} to obtain $\lceil 100/k \rceil$ distinct NTFE FRSTs of each $N-$favorable polytope.

In the Tree ensemble, we took a Delaunay triangulation of each of the two $h^{1,1} = 35$ three-dimensional reflexive polytopes. Then for each polytope, in each $h^{1,1}$ bin with $h^{1,1} \leq 2591$, we took 50 elements of the Tree ensemble, sampled uniformly at random. We also took a further set of 100 trees, 50 from each base polytope, with $h^{1,1} = 2591$, sampled uniformly at random.

In the Skeleton ensemble, we carried out $50{,}000$ Monte Carlo runs starting from $\mathbb{P}^3$, attaining a maximum $h^{1,1}$ of 15{,}970. In each $h^{1,1}$ bin we took 100 Monte Carlo runs uniformly at random from the set of all Monte Carlo runs that reached a base with $h^{1,1}$ in the bin. For each sampled run, from the collection of all bases in that run that fell within the $h^{1,1}$ bin, we selected one base uniformly at random.\footnote{In particular, this means that in the $h^{1,1} = 1$ bin, $\mathbb{P}^3$ was sampled 100 times.}

In the K2 ensemble, we took the toric base $B_3$ described in \S2 of \cite{Yinan}.

\FloatBarrier

\section{Results}\label{sec:results} 

\begin{figure}[htbp]
    \centering
    \includegraphics[width=\linewidth]{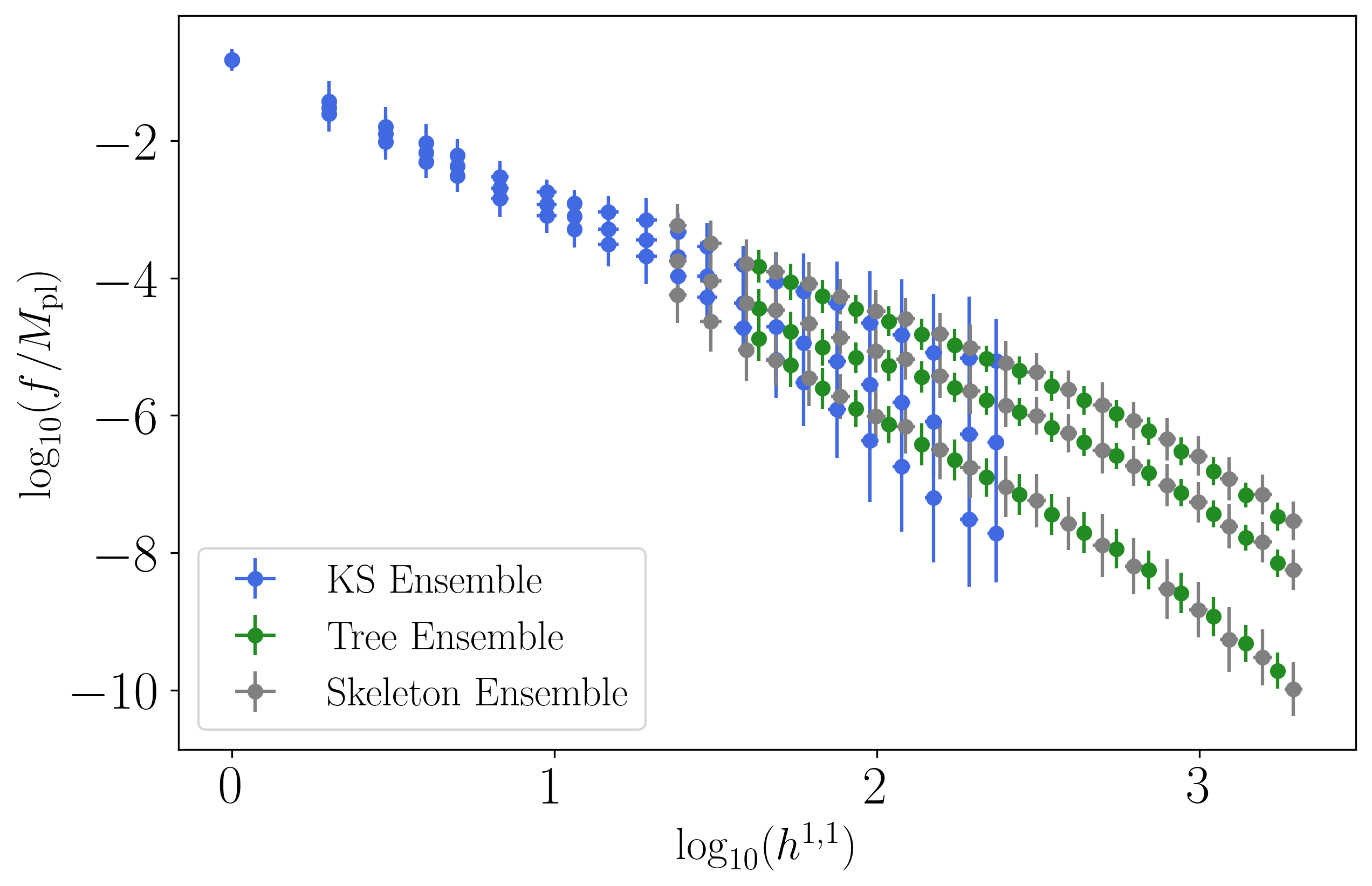} 

    \caption{The mean values across different geometries of $\text{max} \log_{10}(f/M_\text{pl})$, $\text{median} \log_{10}(f/M_\text{pl})$, and $\text{min} \log_{10}(f/M_\text{pl})$ in  the Kreuzer-Skarke, Tree, and Skeleton ensembles.  
    The error bars show the standard deviation in each bin.}
    \label{fig:f_h11}

\end{figure}

\begin{figure}[htbp]
    \centering
    \includegraphics[width=\linewidth]{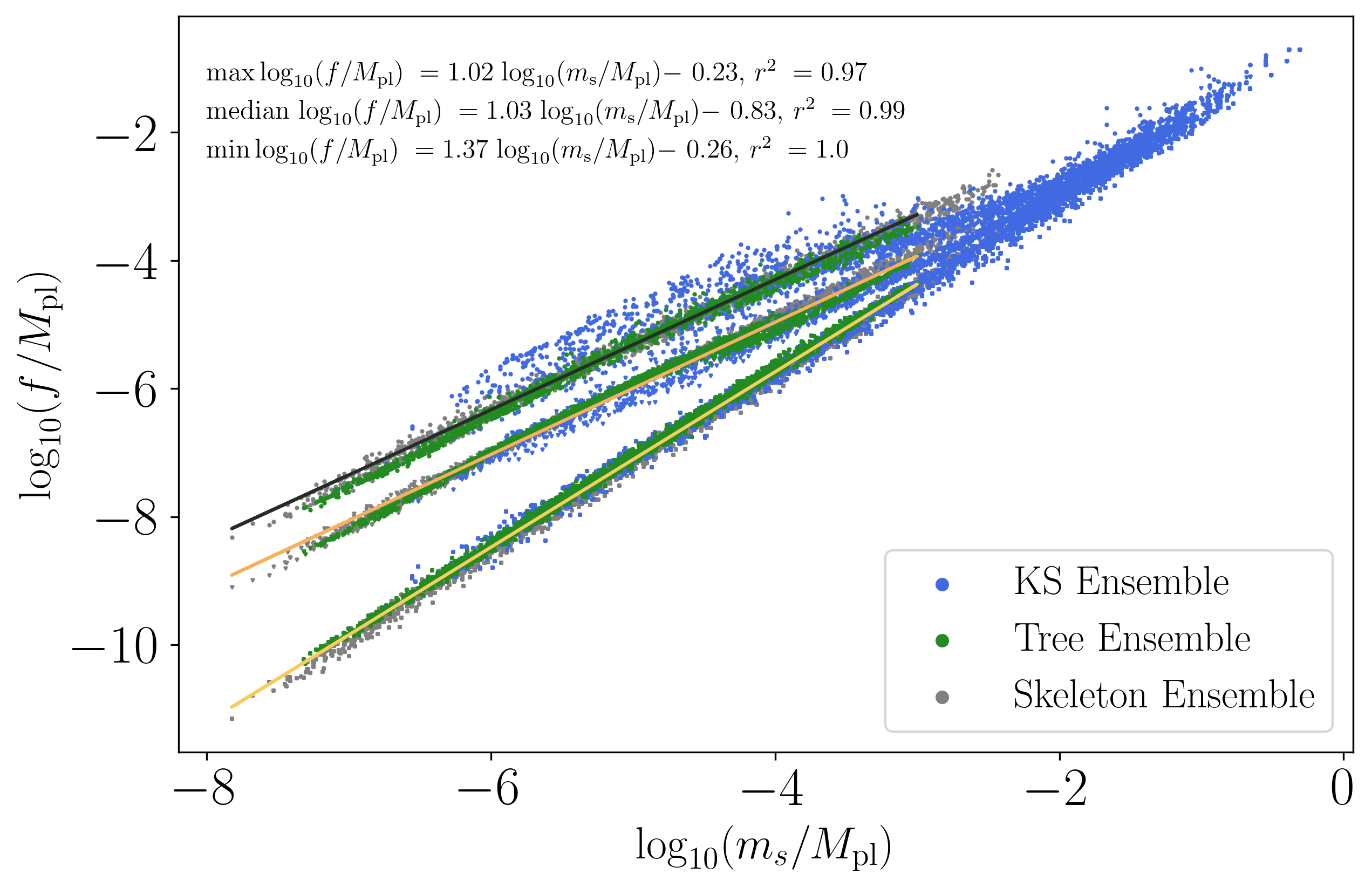} 

    \caption{Decay constants $f$ and the string scale $m_s$. 
    Max, median, and min values are shown.
    The fits shown are taken to the data points with $\log_{10}(m_s/M_\text{pl}) \leq -3$.}
    \label{fig:f_m_string}

\end{figure}

The primary results of this paper are the decay constants,  masses, and couplings to photons of the axions in the Kreuzer-Skarke, Tree, Skeleton, and K2 ensembles: see
Figures \ref{fig:geff_h11}, \ref{fig:f_h11}, \ref{fig:f_m_string}, \ref{fig:m},   \ref{fig:xray_geff}, and \ref{fig:geff_ms}.

\subsection{Decay Constants and Masses}

The axion  decay constants are shown in Figure  \ref{fig:f_h11}.
We see that in each of the ensembles, the decay constants decrease as $h^{1,1}$ increases, with an approximate power-law falloff.
If one fits the data in Figure \ref{fig:f_h11} at $\log_{10}(h^{1,1}) > 1.4,$ one obtains the fits
\begin{align}
\max(f/M_\text{pl}) =& -2.13 \log_{10}(h^{1,1}) -0.26 \qquad\text{with}\qquad\chi^2 = 10.30\,,\\ 
\text{median}(f/M_\text{pl}) =& -2.16 \log_{10}(h^{1,1}) -0.82 \qquad\text{with}\qquad\chi^2 = 15.18\,,\\
\text{min}(f/M_\text{pl}) =&  -2.84  \log_{10}(h^{1,1}) -0.33 \qquad\text{with}\qquad\chi^2 = 6.66\,,
\end{align} each of these with 44 degrees of freedom.  
The above trends are consistent with the results obtained by \cite{Demirtas:2018akl} in the Kreuzer-Skarke case.  
However, the facts that we find comparable trends in two
other classes of string geometries, and that these trends extend to much higher $h^{1,1}$, are striking evidence of universality.

\begin{figure} 
    \centering
    \includegraphics[width=\linewidth]{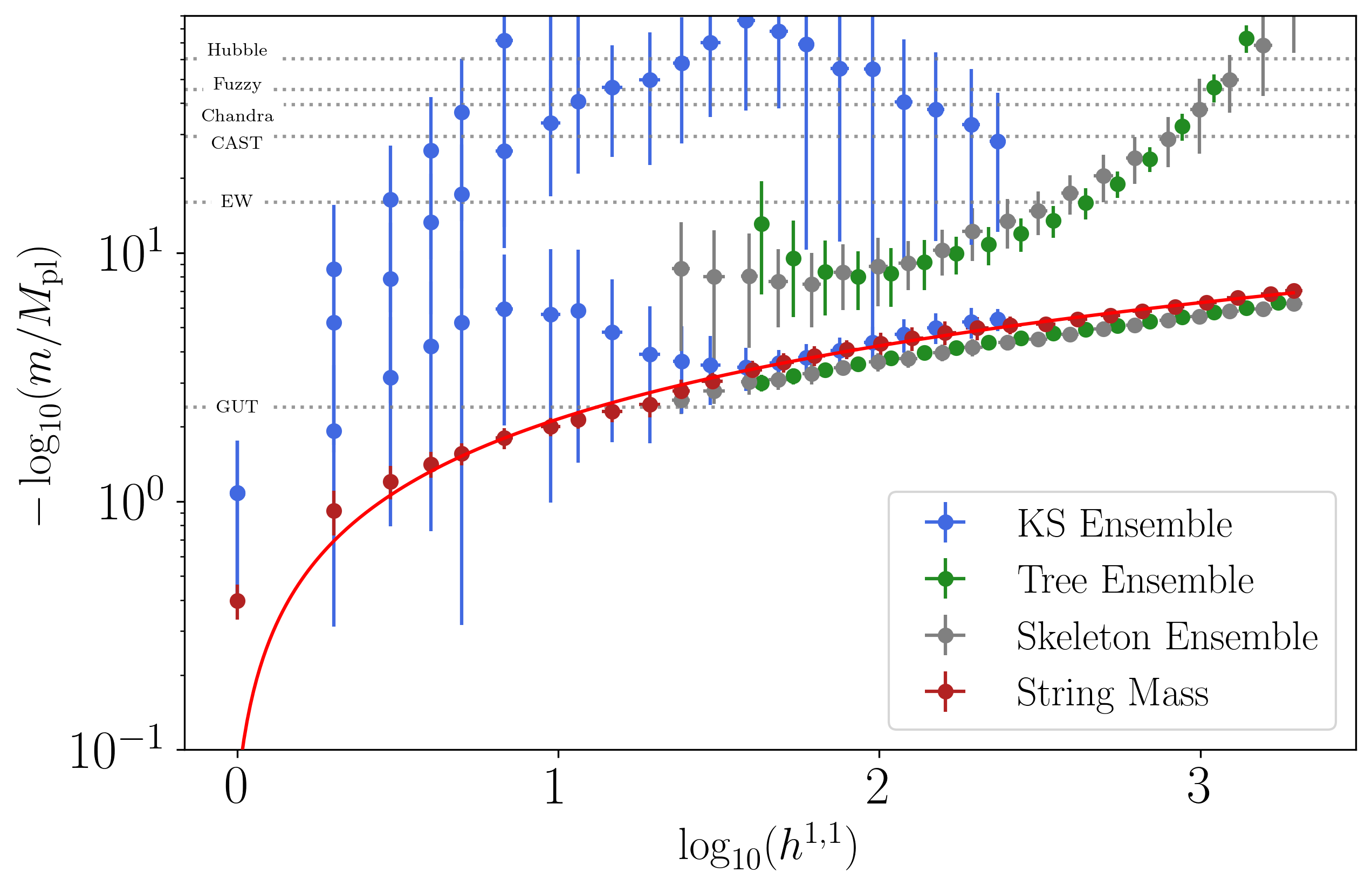} 

    \caption{Axion masses as functions of $h^{1,1}$.  The blue, green, and gray points are the mean axion masses in the Kreuzer-Skarke, Tree, and Skeleton ensembles, respectively, and the error bars are the standard deviation in each bin. 
    The red points show the string mass $m_s$, and the fit in red to $\log_{10}(m_s/M_\text{pl})$ is given by $\log_{10}(m_s/M_\text{pl}) = -2.08 \log_{10}(h^{1,1}) - 0.07$ with $r^2 = 0.94$.   Characteristic energy scales are labeled for reference.}  
    \label{fig:m}
 \end{figure}

The axion masses  are shown in Figure \ref{fig:m}. 
A key finding is that the typical axion masses in our ensembles populate the entire range of interest for observations, from the electroweak scale to the Hubble scale. From these masses and the above decay constant fits, one may estimate the relic abundance of axions produced by misalignment. An initial study of individual species suggests that they are a subleading component of the dark matter. However, a systematic study of the total axion relic abundance in these models is warranted in the future, including potential instability due to decays into dark gluons.

\subsubsection*{Scaling Estimates for  Decay Constants}

In Figure \ref{fig:f_m_string}  one sees that
the maximum, median, and minimum $\log_{10}(f/M_\text{pl})$ 
are very well-approximated as \emph{linear} functions of 
$\log_{10}(m_s/M_\text{pl}),$ where the string mass $m_s$ is given by \begin{equation}m_s = \frac{g_s}{\sqrt{4 \pi \mathcal{V}}} M_\text{pl}\,,\label{eq:string_scale}\end{equation} 
where we have set the string coupling $g_s$ to unity, and $\mathcal{V}$ is the total unwarped volume of the threefold.\footnote{See~\cite{ValeixoBento:2025emh} for a convenient summary of frames and conventions in string compactifications.} 

We now explain that the slopes can be recovered from a simple model, following \cite{Cheng:2025ggf}. 
Equation (3.16) of \cite{Cheng:2025ggf} models the decay constant $f_i$ of the $i$th most massive axion as \begin{equation}f_i \approx \frac{1}{\tau_\text{max}^{3/4} \tau_i^{1/4}}\,,\label{eq:cheng_gendler_fs}\end{equation} where $\tau_\text{max}$ is the maximum volume of a  prime toric divisor, and $\tau_i$ is the $i$th smallest prime toric divisor volume. If we take \begin{equation}\mathcal{V} \approx \tau_\text{max}^{3/2}\,,\end{equation}  
and use equation \eqref{eq:string_scale}, we have 
\begin{equation}\label{eq:fmodel}
f_i \approx \frac{\sqrt{4 \pi}}{g_s}m_s \frac{1}{\tau_i^{1/4}}\,.
\end{equation} 

We find that the minimum divisor volume in the Tree ensemble is typically 1, the median divisor volume is typically much smaller than $\mathcal{V}^{2/3}$, and the maximum divisor volume is approximately $\mathcal{V}^{2/3},$  
so we expect slopes of $1, 1,$ and $4/3$ for the plots of $\max \log_{10}(f/M_\text{pl})$, $\text{median} \log_{10}(f/M_\text{pl})$, and $\min \log_{10}(f/M_\text{pl})$ as functions of $\log_{10}(m_s/M_\text{pl}).$ This is consistent with the fits from Figure \ref{fig:f_m_string}, which give slopes of 1.02, 1.03, and 1.37, respectively.

\subsection{Axion-Photon Couplings}

\begin{figure}[htbp]
\vspace{-3cm}
    \centering
    \includegraphics[width=\linewidth]{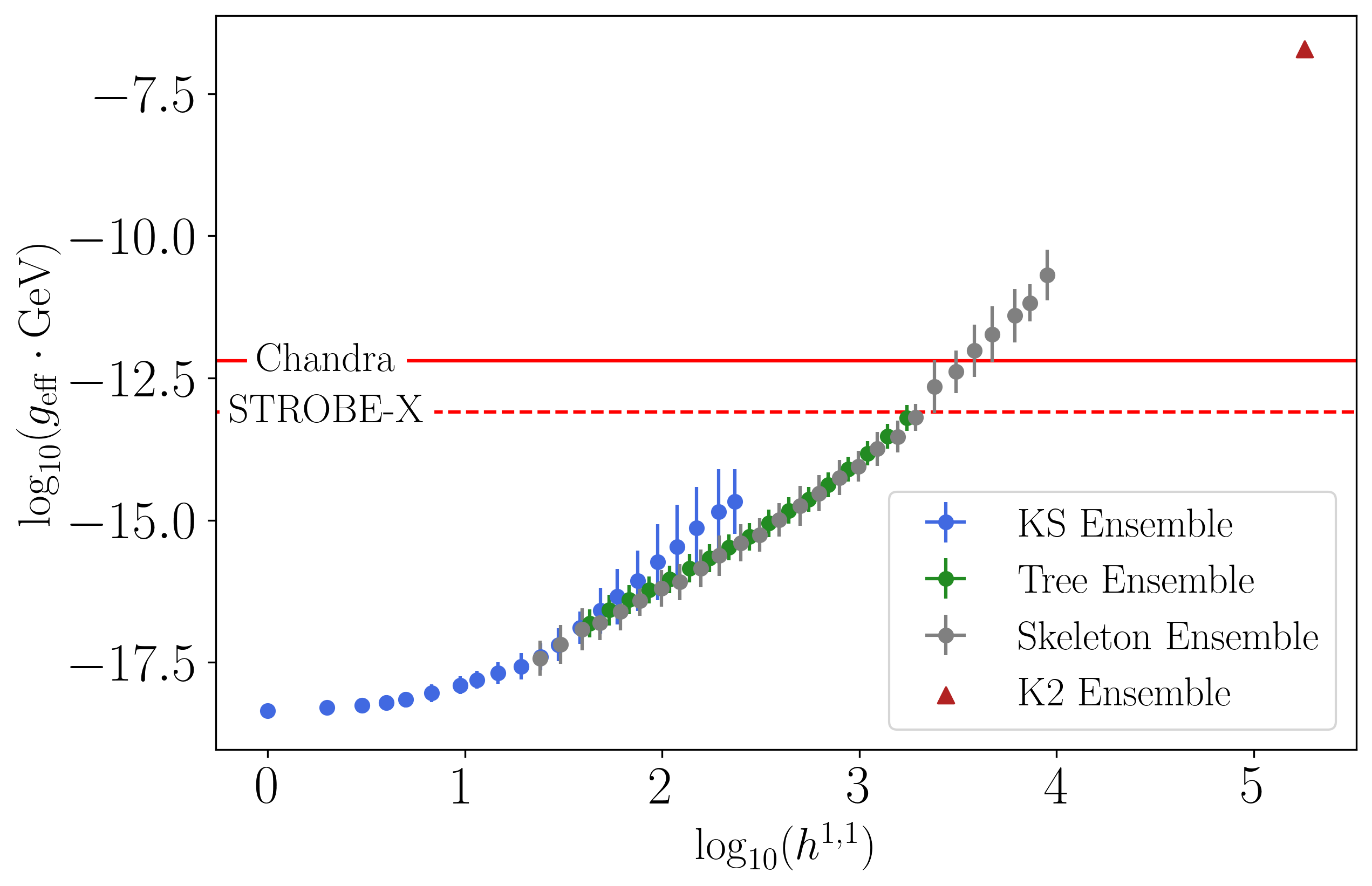} 
    \vspace{-0.8cm}
    \caption{X-ray constraints on the effective axion-photon coupling $g_{\text{eff}}$.
    The error bars show the standard deviation in
    each bin.  Models above the Chandra line are excluded, and models above the projected STROBE-X line  could be tested or excluded. We have imposed
$\alpha^{-1}_\text{EM, UV} \geq 29.5$, and applied a mass cut $m \le
10^{-12} \text{ eV}$.}
    \label{fig:xray_geff}

  \vspace{0.5cm}
    \includegraphics[width=\linewidth]{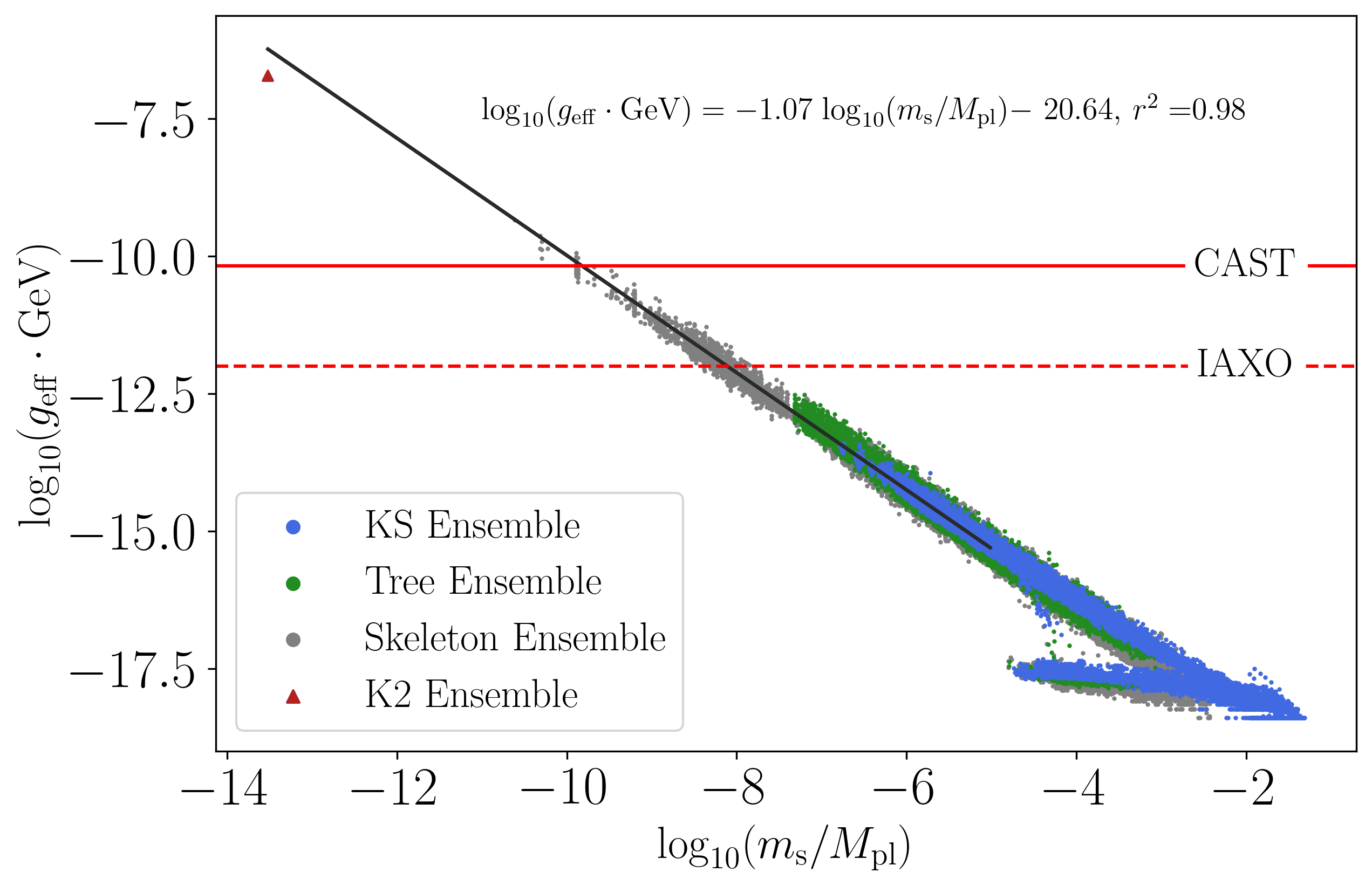} 
    \vspace{-0.5cm}
    \caption{Effective coupling $g_\text{eff}$ as a function of the string scale $m_s$. 
We have imposed $\alpha^{-1}_\text{EM, UV} \geq 19.5$, and a mass cut $m \le
10^{-2} \text{ eV}$.  The fit is to data with $\log_{10}(m_s / M_\text{pl}) \leq -5$.}
    \label{fig:geff_ms}

\end{figure}

One of our principal results is that models in our ensemble with sufficiently high $h^{1,1}$ are in tension with existing helioscope and X-ray limits.  The helioscope results were already shown in Figure   \ref{fig:geff_h11}, and the X-ray results appear in Figure \ref{fig:xray_geff}.

To discuss the resulting exclusions, we first explain the cuts we imposed to compare to these two classes of experiments.  Helioscope experiments such as CAST can probe  axions with $m \lesssim 10^{-2} \text{ eV}$, 
so we imposed a mass cut 
$m \lesssim 10^{-2} \text{ eV}$ when evaluting the sum in \eqref{eq:geff} for the purpose of obtaining the helioscope constraints in Figure \ref{fig:geff_h11}.  On the other hand,
the Chandra X-ray spectrum measurements are sensitive to axions with $m \lesssim 10^{-12} \text{ eV}$, and we used this lower cutoff when evaluting the sum in \eqref{eq:geff} to obtain the X-ray constraints in Figure \ref{fig:xray_geff}.

For both classes of experiments, axions below the 
light threshold $m_{\text{light}}$ have negligible couplings to photons,
and so constraints come from axions with masses $m$ in the range
\begin{equation}\label{eq:mx}
    m_{\text{light}} \le m \le m_{\text{X-ray}}\simeq 10^{-12}\,\text{eV} \,,
\end{equation}
for X-rays, and
\begin{equation}\label{eq:mh}
    m_{\text{light}} \le m \le m_{\text{helioscope}}\simeq 10^{-2}\,\text{eV} \,,
\end{equation}
for helioscopes.

Whether the mass windows given in \eqref{eq:mx} and \eqref{eq:mh} are \emph{nonempty}, i.e., whether they  contain axions that can be tested by the corresponding experiments, depends on the assumptions we make about the ultraviolet completion of electromagnetism. 
A rather conservative assumption is that $\alpha^{-1}_\text{EM, UV} \gtrsim 20$: achieving a smaller value of $\alpha^{-1}_\text{EM, UV}$ would require  the existence of significant amounts of electromagnetically charged matter beyond the SM and the MSSM, which have $\alpha^{-1}_{\text{UV}}\sim 40$ and $\alpha^{-1}_{\text{UV}}\sim 25$, respectively.  
Moreover, for $\alpha^{-1}_\text{EM, UV} \gtrsim 20$, we find that the helioscope window \eqref{eq:mh} is nonempty, and one can set constraints.
Specifically, for the helioscope case we took $\alpha^{-1}_\text{EM, UV} \ge 19.5$.

On the other hand, some models with $20 \le \alpha^{-1}_\text{EM, UV} \le 30$ have $ m_{\text{light}} \ge m_{\text{X-ray}}$, i.e., the X-ray window \eqref{eq:mx} is empty, and hence these models are not constrained by X-rays.
We have not quantified the X-ray limits on models in our ensemble with $19.5 \le \alpha^{-1}_\text{EM, UV} \le 29.5$. Instead, we \emph{only state the X-ray constraints} that apply to models satisfying the stronger
assumption $\alpha^{-1}_\text{EM, UV} \ge 29.5$, which still requires significant electromagnetically charged matter beyond the SM.
Said differently, the X-ray limits on our ensemble are more sensitive to ultraviolet assumptions than the helioscope limits are.
 
In our ensemble, the effective axion-photon coupling  $g_{\text{eff}}$ defined in \eqref{eq:geff} increases monotonically with $h^{1,1}$.
This was already shown for the helioscope case in 
Figure \ref{fig:geff_h11}, but is likewise evident for the X-ray case
shown in Figure \ref{fig:xray_geff}.  
We find that models with $h^{1,1} \gtrsim 5{,}000$ are in tension with the \emph{current} constraints from Chandra.
This result can be compared with the helioscope case, where models with 
$h^{1,1} \gtrsim 10{,}000$ are in tension with the current constraints from CAST, and models with 
$h^{1,1} \gtrsim 5{,}000$ could be probed by future helioscopes such as IAXO.\footnote{We reiterate that the X-ray analysis in our ensemble was performed with more restrictive assumptions about $\alpha^{-1}_\text{EM, UV}$ than we used for the helioscope analysis.}

Our results differ from the  projected value of $g_\text{eff}$ obtained in the earlier work \cite{Halverson:2019cmy} for $h^{1,1}=2{,}483$ in the Tree Ensemble: $g_{\text{eff}}^{\text{here}} \sim g_{\text{eff}}^{\text{there}}/20$. 
This can be attributed  
to a number of differences in method in \cite{Halverson:2019cmy}, including a less precise mass model, the absence of suppression effects from the QED instanton, and the projection from a power-law scaling from computations at $h^{11}\lesssim 250$. In contrast, the computational breakthroughs of the present work
have allowed us not just to perform the calculation at $h^{1,1}=2{,}483$, but also to move on to far larger values of $h^{1,1}$. 

\subsubsection*{Scaling Estimates for Axion-Photon Couplings}

In Figure \ref{fig:geff_ms}, we see that the axion-photon couplings are closely correlated with the string scale $m_s$, as expected from the $m_s$-$f$ correlation seen in Figure \ref{fig:f_m_string}.
We now provide a scaling estimate supporting the observed $g_{\text{eff}}$-$m_s$ correlation.

Equation \eqref{eq:em_coupling_for_later}
gives the axion-photon coupling for the $a$th mass-eigenstate axion as \begin{equation}g_{a\gamma\gamma} = \frac{1}{137}  \sum_{b} Q^\text{EM}_{ b} (LH)_{ba}\,.\label{eq:hierarchical_geff_from_here}\end{equation}
If this axion is sufficiently massive to avoid the light-threshold suppression effects of \cite{glimmers}, then the term in the sum in equation \eqref{eq:hierarchical_geff_from_here} will be of order $f_a^{-1}$, and so using equation \eqref{eq:fmodel}, one expects $g_{a\gamma\gamma}\propto m_s^{-1}.$ Hence, for a fixed axion-photon experiment, provided there is an axion between the light threshold and the mass cutoff of the experiment, the effective axion-photon coupling should be proportional to $m_s^{-1}$. This is consistent with the slope in the fit 
to the effective axion-photon coupling shown in Figure \ref{fig:geff_ms}\,,
\begin{equation}\log_{10}(g_\text{eff} \cdot \text{GeV}) = -1.07 \log_{10}(m_s/M_\text{pl}) - 20.64\,.\end{equation}  
One can also fit $g_\text{eff}$  
as a function of $\log_{10}(h^{1,1})$ for $\log_{10}(h^{1,1}) > 1.5$, obtaining \begin{equation}\log_{10}(g_\text{eff} \cdot \text{GeV}) = 2.508 \log_{10}(h^{1,1})-21.371\,,\end{equation} with $\chi^2 = 18$ in 49 degrees of freedom.

\FloatBarrier 

\section{Conclusions}\label{sec:conclusions}
 
We have initiated the systematic study of theories of thousands of axions resulting from F-theory. 
We compactified F-theory to four dimensions on Calabi-Yau fourfolds built as Weierstrass models over toric threefold bases, constructing large ensembles following \cite{trees,skeleton,Yinan}.
In these theories we characterized the couplings of 
axions obtained from reduction of the Ramond-Ramond four-form $C_4$.   

Modeling the nonperturbative superpotential by including contributions from prime toric divisors, and working at the tip of the stretched K\"ahler cone, we computed the axion Lagrangian.   We used a non-Higgsable gauge group with an appropriate value of the gauge coupling as a toy model for electromagnetism, and computed the couplings of axions to this ``photon''.  We believe that the results we obtained with these approximations capture in broad strokes the characteristics of axion couplings in the geometric regime of F-theory.\footnote{Unknowns involving, for example, Pfaffian prefactors and the F-theory K\"ahler potential do affect the precise numerical values of the couplings, but have limited impact on our  distributional results.}  On the other hand, we have not drawn any conclusions about axions in F-theory compactifications with strongly curved threefold bases, or in the presence of fluxes that break the gauge group and give  axions large masses by the St\"uckelberg mechanism.

The computational task of constructing ensembles with as many as 181,200 axions was nontrivial.
The first step, construction of Weierstrass models over toric bases, was relatively straightforward: we used the ideas of
\cite{trees,skeleton,Yinan}, and used a new \texttt{CYTools} implementation of general (potentially not weak Fano) toric varieties that
was generously shared with us by J.~Moritz.   
However, finding the tip of the stretched K\"ahler cone in cases with $h^{1,1} \gtrsim 3{,}000$ required sophisticated optimization methods.  Moreover, the axion mass matrices featured gigantic but irregular hierarchies, and brute force numerical diagonalization with high precision was inaccurate for $h^{1,1} \gtrsim 2{,}000$.  We combined high precision with perturbation theory methods to achieve sufficient accuracy.  The computational tools we have developed will facilitate future work on axions in F-theory. 

One of our main findings is that in our ensembles of theories, the axion decay constants scale as a power-law with $h^{1,1}$: see Figure \ref{fig:f_h11}.  Such scaling was discovered in type IIB compactifications on Calabi-Yau threefold hypersurfaces in \cite{Demirtas:2018akl}, for $h^{1,1} \le 491$, and here we have shown that the trends continue up to much larger values of $h^{1,1}$.
The scaling behavior is driven by the empirical fact that the volume of a compactification evaluated at the tip of the stretched K\"ahler cone grows with $h^{1,1}$.  Thus, the trend we find is a property of weakly-curved compactifications,
and understanding whether similar patterns arise for strongly-curved threefold bases is an important open question.

Similarly, the axion masses decrease as $h^{1,1}$ increases. 
The maximum 
mass tracks closely with the string scale: see Figure \ref{fig:m}.
The median axion mass tends to reside in the range relevant for cosmology, while the lightest axion quickly becomes effectively massless.

Another key result is that the axion-photon couplings in our ensemble grow with $h^{1,1}$: see Figure \ref{fig:geff_h11}. 
The axion-photon coupling is very closely correlated with the volume, and hence with the string scale.  
Models in our ensemble with $h^{1,1} \gtrsim 10{,}000$ are in tension with limits on axion-photon couplings from CAST \cite{CAST:2017uph}.  This finding reproduces the results of \cite{Halverson:2019cmy}, which were obtained by direct computation in the Tree ensemble of \cite{trees} up to $h^{1,1}=250$, and were extrapolated to higher $h^{1,1}$.
Here we have carried out the explicit computation 
in the Tree ensemble up to $h^{1,1}=1{,}753$, in the Skeleton ensemble up to $h^{1,1}=9{,}336$, and in the K2 ensemble at $h^{1,1}=181{,}200$.

We conclude that, among F-theory compactifications with weakly-curved threefold bases from the ensembles of \cite{trees,skeleton,Yinan}, and
with visible sectors on seven-branes,
\emph{those with more than 10{,}000 axions are disfavored by experiment under the model assumptions we have made.}
This finding is provisional because we have used a proxy for the photon, and to a somewhat lesser extent because of unknowns related to our model of the superpotential and K\"ahler potential. 

Another effect that could modify our findings is the inclusion of fluxes that induce axion masses by the St\"uckelberg mechanism.  In such a case the effective number of axions could be significantly smaller
than the $h^{1,1}(B)$ obtained in the absence of fluxes: perhaps of order $N_\text{eff} \sim h^{1,1}(B)/2$, 
cf.~\eqref{eq:stuck}.
A systematic study of St\"uckelberg masses in F-theory models with many axions is an important task,
but is well beyond the scope of this work.  

This paper is merely a first step in exploring the F-theory axiverse, and there are many natural avenues for future work. The extremely rich spectra of axions and hidden-sector states in models with large $h^{1,1}$ could lead to a correspondingly complex cosmological history.  It would be worthwhile to understand axion reheating \cite{Halverson:2019kna}, glueball production \cite{glueballs}, 
and stasis \cite{Dienes:2021woi,Dienes:2022zgd,Dienes:2023ziv} (which allows a potential axiverse realization \cite{Halverson:2024oir})
in this setting.  
Particularly in view of the vast numbers of axions and hidden sectors at hand, and the prospect of trends with $h^{1,1}$ analogous to what we have established for 
$g_{a\gamma\gamma}$, we expect that such cosmological phenomena could yield signals at upcoming experiments or provide strong constraints on the theory.
Axion cosmology in F-theory thus presents
an excellent opportunity to connect explicit large-scale string constructions to upcoming cosmological probes. 
The requisite calculations are intricate,  
but  
we have now 
demonstrated 
the existence of computational tools equal to the task.

Axions might also have conceptual implications for the string landscape through their impact on 
counting vacua that are consistent with observations.  
If one supposes that the landscape of four-dimensional vacua of F-theory is finite, a natural question is which sorts of vacua dominate the counting measure on this finite set.  The single F-theory compactification with the most flux vacua \cite{Taylor:2015xtz} has a modest number of axions, so at first sight there is no connection to our study of theories with vast numbers of axions.  However, we quote an observation made in \cite{Yinan}:
there are $\mathcal{O}(10^{45,000})$ potentially inequivalent toric threefolds supporting Weierstrass models with $h^{1,1}=181{,}200$, and each of these allows many flux vacua. It is not inconceivable that, counting the choices of geometry and flux together, a typical F-theory compactification is one with $\mathcal{O}(10^5)$ axions.  If this were found to be true, then our results on axion-photon couplings would take on new meaning: they would imply that axion-photon experiments already exclude typical (in the counting measure) F-theory compactifications in the geometric regime!  We stress that such a hypothetical conclusion would not be experimental evidence against F-theory itself: one could view the experiments either as refuting the null hypothesis that observables can be computed using the counting measure on the landscape of vacua, or as excluding weakly-curved solutions.

In conclusion, experimental limits on axion-photon couplings have the potential to exclude swaths of the F-theory landscape. Opening this window on fundamental physics will require advances in constructing visible sectors and stabilizing moduli in F-theory, and 
could lead to a compelling  
new phase
in the study of F-theory.

\section*{Acknowledgements}

We thank Alexander Bellas for collaboration in the early stages of this work.
We thank Josh Benabou, Federico Compagnin, Naomi Gendler, Thomas Harvey, Manki Kim, Cody Long, Nate MacFadden, M.C.~David Marsh, David J.E.~Marsh, Jakob Moritz, Richard Nally, 
Arthur Platschorre, Fabian Ruehle, Ben Safdi,
Elijah Sheridan, Washington Taylor, and Yi-Nan Wang for discussions. S.V.P.F. is particularly grateful to Elijah Sheridan for extensive guidance. This work was completed in part using the Explorer Cluster, supported by Northeastern’s University's Research Computing team.  S.V.P.F.~acknowledges travel support from the Leverhulme Trust under Grant No.~RPG-2022-145.
The research of L.M.~and S.V.P.F.~is supported in part by NSF grant PHY-2309456.  J.H. is supported in part  by NSF grant PHY-2209903.

\newpage 

\begin{appendix}

\addtocontents{toc}{\protect\setcounter{tocdepth}{1}}

\addtocontents{toc}{\protect\setcounter{tocdepth}{1}}
\section{Details of the Ensembles}\label{sec:appensembles}

To describe four-dimensional physics, one compactifies F-theory on a smooth Calabi-Yau fourfold $\hat{Y}_4$, equipped with a surjective morphism $\pi : \hat{Y}_4 \rightarrow B_3$ for some compact K\"ahler threefold $B_3$, such that $\pi^{-1}(p)$ is a smooth elliptic curve for a generic $p\in B_3$. 
A standard approach is to first construct a toric threefold base $B_3,$ and then consider the Calabi-Yau fourfold $Y_4$ defined by the Weierstrass equation 
\begin{equation}y^3 = x^3 + fxz^4 + g z^6\,,
\end{equation} 
where $f \in \mathcal{O}_{B_3}(-4K_{B_3})$, $g \in \mathcal{O}_{B_3}(-6 K_{B_3}),$ and where $K_{B_3}$ is the canonical class of $B_3.$ Such a $Y_4$ is Calabi-Yau, but may be singular. When $Y_4$ has codimension one singularities (in the base) there exists a resolution $\hat{Y}_4 \rightarrow Y_4$ that resolves them. Such a resolution $\hat{Y}_4$ is in general not unique and corresponds to moving out on the $d=3$, $\mathcal{N}=2$ M-theory Coulomb branch.  The four-dimensional F-theory physics deduced from the Coulomb branch must be independent of the choice of resolution since one must come off the Coulomb branch to take the F-theory limit. In practice then, to construct an F-theory compactification at this level of detail, one just needs to construct a K\"ahler threefold $B_3$ and find $f \in \mathcal{O}_{B_3}(-4 K_{B_3})$ and $g \in \mathcal{O}_{B_3}(-6 K_{B_3})$ such that $B_3$ satisfies the conditions \eqref{eqn:finite_distance} to be at finite distance in the moduli space.\footnote{A general F-theory compactification also requires a choice of fluxes $G_4$, but here, we assume that the constraint \begin{equation}G_4 + \frac{1}{2}c_2(\hat{Y}_4) \in H_4(\hat{Y}_4, \mathbb{Z})\,,\end{equation} where $c_2(\hat{Y}_4)$ is the second Chern class of $\hat{Y}_4$, can be satisfied by setting $G_4 = 0$ or alternatively that the effects of $G_4$ on gauge sectors and axion masses (via the St\"uckelberg mechanism) are negligible. Relaxing this constraint is an important direction for future work.} 

Two ensembles of such K\"ahler threefold bases are the Tree ensemble \cite{trees} and the Skeleton ensemble \cite{skeleton}. Both of these ensembles consist of large numbers of toric threefold bases $B_3$ constructed as blowups of simpler toric threefolds. For toric threefolds $B_3$, sections of line bundles have an easy description in terms of combinatorial data.  
Let $\Sigma$ be the normal fan associated to $B_3$ and 
let $\Sigma(n)$ denote the set of $n$-dimensional cones of $\Sigma$. Then for each ray $\rho \in \Sigma(1)$, let $u_\rho$ denote its minimal generator. Let $M$ denote the lattice of monomials, and let $N$ denote the dual lattice which houses the toric fan. A global section $s$ of a line bundle $\mathcal{O}_{B_3}(\sum_{\rho \in \Sigma(1)} a_\rho D_\rho)$ can then be written as 
\begin{equation}s = \sum_{m \in \mathcal{S}} a_m \prod_{\rho \in \Sigma(1)}x_\rho^{\langle m, u_\rho \rangle }\,,
\end{equation} 
where $x_{\rho}$ is the homogeneous coordinate associated to $D_\rho,$ $a_m \in \mathbb{C},$ 
and 
\begin{equation}\mathcal{S} = \{m \in M | \langle m, u_\rho \rangle \geq - a_\rho \text{ for all }\rho \in \Sigma(1)\}\,.
\end{equation} 
Hence, defining the lattice polytopes 
\begin{equation}\mathcal{F} := \{m \in M | \langle m, u_\rho\rangle \geq -4 \text{ for all }\rho \in \Sigma(1)\}\,, \label{eq:f_poly_def}
\end{equation} 
and 
\begin{equation}
\mathcal{G} := \{m \in M | \langle m, u_\rho \rangle \geq -6 \text{ for all }\rho \in \Sigma(1)\}\,,\label{eq:g_section_definition}
\end{equation} 
any sections $f \in \mathcal{O}_{B_3}(-4K_{B_3})$ and $g \in \mathcal{O}_{B_3}(-6 K_{B_3})$ can be written as 
\begin{equation}
f = \sum_{m \in \mathcal{F}} a_m\prod_{\rho \in \Sigma(1)} x_\rho^{\langle m, u_\rho \rangle + 4}\,,
\end{equation} 
and 
\begin{equation}
g = \sum_{m \in \mathcal{G}} b_m\prod_{\rho \in \Sigma(1)} x_\rho^{\langle m, u_\rho \rangle + 6}\,,\label{eq:g_section_real_def}
\end{equation} 
for some $a_m, b_m \in \mathbb{C}.$ 

\subsection{Multiplicities on toric subvarieties}

The condition \eqref{eqn:finite_distance} is  necessary  in order to be at finite distance in moduli space, and so we are interested in constructing ensembles of fourfolds that obey  \eqref{eqn:finite_distance}.   
Happily, there exists a simple condition that ensures that a smooth toric threefold $X_\Sigma$ satisfies the conditions in \eqref{eqn:finite_distance}. Let  $\mathcal{P}_{\mathcal{G}} \subseteq \mathbb{R}^3$ be the polytope given by \begin{equation}\mathcal{P}_{\mathcal{G}} = \text{conv}(\mathcal{G})\,,\label{eq:p_definition}\end{equation} where $\mathcal{G}$ is defined in equation \eqref{eq:g_section_definition}. Note that for any $X_\Sigma,$ $(0, 0, 0) \in \mathcal{G}$.  We will make use of the following:
\begin{equation}\label{polytopecondition}
\text{\emph{Origin condition}:}~~\text{$(0, 0, 0)$ is in the strict interior of $\mathcal{P}_{\mathcal{G}}$}\,.
\end{equation}
We have the following proposition that provides a sufficient condition for equation \eqref{eqn:finite_distance} being satisfied on all toric subvarieties:

\newtheorem{proposition}{Proposition}
\begin{proposition}\label{prop:poly}
Suppose $X_\Sigma$ is a smooth toric threefold with fan $\Sigma$, and let $\mathcal{P}_{\mathcal{G}}$ be defined as in equation \eqref{eq:p_definition}. Suppose $(0, 0, 0)$ is in the strict interior of $\mathcal{P}_{\mathcal{G}}$. Then for generic $f \in \mathcal{O}_{X_\Sigma}(-4K)$ and $g \in \mathcal{O}_{X_\Sigma}(-6K)$, where $K$ is the canonical class of $X_\Sigma,$ 
we have that $X_\Sigma$ satisfies equation \eqref{eqn:finite_distance} on all toric subvarieties. 
Specifically, for toric divisors $D$, toric curves $C$, and toric points $p$, we have
\begin{enumerate}
    \item[a.] \text{mult}$_D(f,g) < (4,6)$\,,
    \item[b.] \text{mult}$_C(f,g) < (8,12)$\,, 
    \item[c.] \text{mult}$_p(f,g) < (12,18)$\,.
\end{enumerate}
\end{proposition}

We remark that it was shown in \cite{skeleton} that the assumptions of Proposition \ref{prop:poly} imply condition (a), 
i.e.~that $X_\Sigma$ has no $(4,6)$ divisors. 
Here we address the remaining two conditions, and so also exclude $(8,12)$ curves and $(12,18)$ points.\\

\textit{Proof:} For each ray $\rho \in \Sigma(1)$ in the fan of $X_\Sigma,$ let $u_\rho$ be the corresponding minimal generator. We need to check that all torus-invariant divisors $D$ of $X_\Sigma$ have $\text{mult}_D(f, g) < (4,6)$, that all torus-invariant curves $C$ of $X_\Sigma$ have $\text{mult}_{C}(f, g) < (8, 12),$ and that all torus-invariant points $p$ of $X_\Sigma$ have $\text{mult}_p(f, g) < (12, 18).$ Note that by the orbit-cone correpondence, every torus-invariant subvariety corresponds to a cone in $\Sigma$, and hence the subvariety can be written as the intersection of the prime toric divisors whose rays span the cone: see Theorem 3.2.6 in \cite{CLS}. 

We start with divisors. Note that if $D_{\rho_1}$ is a prime toric divisor with $\text{mult}_{D_{\rho_1}}(f, g) \geq (4, 6)$, then in particular, for all $m' \in \mathcal{G}$ we have 
\begin{align}
6 & \leq \text{mult }_{D_{\rho_1}}(g) \\ & = \min_{m \in \mathcal{G}} \left\{\langle m, u_{\rho_1} \rangle + 6\right\} \\ & \leq \langle m', u_{\rho_1} \rangle + 6\,.
\end{align} 
Hence, since all points $m' \in \mathcal{G}$ satisfy 
\begin{equation}
0 \leq \langle m', u_{\rho_1} \rangle\,,\label{eq:four_six_hyperplane}
\end{equation} equation \eqref{eq:four_six_hyperplane} provides a bounding hyperplane for $\mathcal{P}_{\mathcal{G}}.$ Since $m' = (0, 0, 0)$ saturates the hyperplane inequality, $(0, 0, 0)$ is not in the strict interior of $\mathcal{P}_{\mathcal{G}}$, a contradiction. Hence, $X_\Sigma$ cannot have any (4, 6) divisors.

Moving on to torus-invariant curves $C$, we know that $C$ can be written as the intersection of two prime toric divisors, so without loss of generality, we can write 
\begin{equation}C = D_{\rho_1} \cap D_{\rho_2}\,.
\end{equation} 
Suppose also that $\text{mult}_{C}(f, g) \geq (8, 12).$ Then in particular, for all $m' \in \mathcal{G}$ we have 
\begin{align}
12 & \leq \text{mult}_C(g) \\ & = \min_{m \in \mathcal{G}} \left\{(\langle m, u_{\rho_1} \rangle + 6 ) + (\langle m, u_{\rho_2}\rangle + 6)\right\} \\ & \leq \langle m', u_{\rho_1} + u_{\rho_2} \rangle + 12\,.
\end{align} 
Hence, all points $m' \in \mathcal{G}$ satisfy 
\begin{equation}
0 \leq \langle m', u_{\rho_1} + u_{\rho_2} \rangle\,.
\end{equation} 
This provides a hyperplane constraint for $\mathcal{P}_{\mathcal{G}}$ which is saturated for $m' = (0, 0, 0).$ Hence, $(0, 0, 0)$ is not in the strict interior of $\mathcal{P}_{\mathcal{G}},$ a contradiction. Hence, $X_\Sigma$ cannot have any (8, 12) curves.

Lastly, suppose $p$ is a torus-invariant point. Then without loss of generality we can write $p = D_{\rho_1} \cap D_{\rho_2} \cap D_{\rho_3}$. Suppose also that $p$ has $\text{mult}_p(f,g) \geq (12, 18).$ Then in particular, for all $m' \in \mathcal{G},$ we have 
\begin{align}
18 & \leq \text{mult}_p(f,g) \\ & = \text{min}_{m \in \mathcal{G}} \left\{(\langle m, u_{\rho_1} \rangle + 6) + (\langle m, u_{\rho_2} \rangle + 6) + (\langle m, u_{\rho_3} \rangle + 6)\right\} \\ & \leq \langle m', u_{\rho_1} + u_{\rho_2} + u_{\rho_3} \rangle + 18\,.
\end{align}  
Hence, all points $m' \in \mathcal{G}$ satisfy 
\begin{equation}
0 \leq \langle m', u_{\rho_1} + u_{\rho_2} + u_{\rho_3} \rangle\,.
\end{equation} 
This provides a hyperplane constraint for $\mathcal{P}_{\mathcal{G}}$ which is saturated for $m' = (0, 0, 0).$ Hence, $(0, 0, 0)$ is not in the strict interior of $\mathcal{P}_{\mathcal{G}},$ a contradiction. Hence, $X_{\Sigma}$ cannot have any $(12, 18)$ points. $\square$

We have the following proposition concerning the final base reached in any Monte Carlo chain.

\begin{proposition}\label{prop:finalbase}
Suppose $X_{\Sigma}$ is a toric variety satisfying Proposition $1$, such that any toric blowup of $X_{\Sigma}$ yields a toric variety $X_{\Sigma'}$ with corresponding $\mathcal{P}_{\mathcal{G}'}$ \textbf{not} containing $(0, 0, 0)$ in its strict interior (i.e. all $X_{\Sigma'}$ violate condition \eqref{polytopecondition}), and let $K$ be the canonical class of $X_{\Sigma}.$ Then for generic choices of $f \in \mathcal{O}_{X_{\Sigma}}(-4K)$ and $g \in \mathcal{O}_{X_{\Sigma}}(-6K)$, 
\begin{enumerate}
    \item[a.] all toric curves $C$  have $\text{mult}_C(f, g) < (4,6)$, and
    \item[b.] all toric points $p$ have $\text{mult}_p(f, g) < (8, 12)$,
\end{enumerate} 
i.e., the multiciplity of vanishing is significantly lower than required by \eqref{eqn:finite_distance}.
\end{proposition}

In \cite{skeleton} it was proved
that with the assumptions of Proposition \ref{prop:finalbase},
the conclusion (a) follows. In particular, the conditions of Proposition \ref{prop:finalbase} are fulfilled in all final bases reached in the Monte Carlo of \cite{skeleton}. The condition (b) was observed in \cite{skeleton} as an empirical result, and here we complete the proof.

\textit{Proof:} We proceed by contradiction, in each case producing a toric blowup of $X_{\Sigma}$ yielding a smooth compact $X_{\Sigma'}$ with $(0, 0, 0)$ in the strict interior of $\mathcal{P}_{\mathcal{G}'}.$ 

To standardize notation, let $\rho_1, \ldots, \rho_{N+3} \in \Sigma(1)$ be the rays corresponding to the prime torics of $X_{\Sigma},$ denoted $D_{\rho_1}, \ldots, D_{\rho_{N+3}}.$ Let the minimal generator of ray $\rho_i$ be denoted $u_{\rho_i}.$ 

Suppose by way of contradiction that there exists a toric curve $C \subseteq X_{\Sigma}$ such that for generic choices of $f, g$, $\text{mult}_C(f, g) \geq (4, 6).$ Since $C$ is torus invariant, it can be written as the intersection of two prime toric divisors. Without loss of generality, we can assume $C = D_{\rho_1} \cap D_{\rho_2}$. Then since $\text{mult}_C(f, g) \geq (4, 6),$ in particular, we have that for any $m' \in \mathcal{G},$ 
\begin{align}
6 & \leq \text{mult}_C(g) \\ & = \min_{m \in \mathcal{G}}\left\{(\langle m, u_{\rho_1} \rangle + 6) + (\langle m, u_{\rho_2} \rangle + 6) \right\} \\ & \leq \langle m', u_{\rho_1} + u_{\rho_2} \rangle + 12\,. 
\end{align} 
Rearranging, this means that for all $m' \in \mathcal{G},$ we have 
\begin{equation}
\langle m', u_{\rho_1} + u_{\rho_2} \rangle \geq -6\,. 
\end{equation} 
That is, one can perform a toric blowup of $C$ without changing $\mathcal{G}.$ In other words, the toric variety $X_{\Sigma'}$ obtained by blowing up $C$ has $(0, 0, 0)$ in the interior of $\mathcal{P}_{\mathcal{G}'}.$ This contradicts the fact that $X_{\Sigma}$ was the final base in a Monte Carlo run.

Similarly, suppose by way of contradiction that there exists a toric point $p \in X_\Sigma$ such that for generic choices of $f, g$, $\text{mult}_p(f, g) \geq (8, 12).$ Since $X_{\Sigma}$ is smooth, the maximal cone corresponding to $p$ is generated by three rays. Hence, $p$ can be written as the intersection of three prime toric divisors. Without loss of generality, we can assume $p = D_{\rho_1} \cap D_{\rho_2} \cap D_{\rho_3}.$ Then since $\text{mult}_p(f, g) \geq (8, 12)$, in particular, we have that for any $m' \in \mathcal{G},$ \begin{align}12 & \leq \text{mult}_p(g) \\ & = \text{min}_{m \in \mathcal{G}} \left\{(\langle m, u_{\rho_1} \rangle + 6) + (\langle m, u_{\rho_2} \rangle + 6) + (\langle m, u_{\rho_3} \rangle + 6)\right\} \\ & \leq \langle m', u_{\rho_1} + u_{\rho_2} + u_{\rho_3} \rangle + 18\,.\end{align} Rearranging, this means that for all $m' \in \mathcal{G}$, we have \begin{equation}\langle m', u_{\rho_1} + u_{\rho_2} + u_{\rho_3} \rangle \geq -6\,.\end{equation} That is, one can perform a toric blowup at $p$ without changing $\mathcal{G}.$ In other words, the toric variety $X_{\Sigma'}$ obtained by blowing up at $p$ has $(0, 0, 0)$ in the interior of $\mathcal{P}_{\mathcal{G}'}.$ This contradicts the fact that $X_{\Sigma'}$ was the final base in a Monte Carlo run. $\square$

We note that even in the absence of toric $(4, 6)$ curves, there may be non-toric $(4, 6)$ curves: see \cite{Yinan}.

\subsection{The Ensembles}

The Tree \cite{trees}, Skeleton, and K2 ensembles 
are designed, in part, in order to fulfill the criterion \eqref{eqn:finite_distance},
which is sufficient to be at finite distance in moduli space.
The Tree ensemble is constructed so that if $\Delta^\circ$ is a three-dimensional polytope whose triangulation is used as the starting point to construct a tree geometry $X_{\Sigma}$, then the height six condition satisfied by all tree geometries (see Section \ref{sec:tree}) implies that the polar dual of $\Delta^\circ$, $\Delta$, is contained in $\mathcal{P}_{\mathcal{G}}.$  Since $\Delta$ contains the origin in its strict interior, so does $\mathcal{P}_{\mathcal{G}}$ and therefore the assumptions of Proposition \ref{prop:poly} are fulfilled. In conclusion, any tree built on $\Delta^\circ$ lacks $(4,6)$ divisors, $(8,12)$ curves, and $(12,18)$ points; this is the original result of \cite{trees}, but we have cast it into the light of our more general theorem. In the Skeleton ensemble, at each step of the Monte Carlo, one only performs a blowup leading to a base that contains $(0, 0, 0)$ in the strict interior of $\mathcal{P}_{\mathcal{G}}$. The K2 ensemble is governed by similar logic to the Tree ensemble, as explained in \S\ref{sec:k2ens}.

\subsubsection{Tree Ensemble}
To describe the Tree ensemble, we first standardize some notation. Let $T$ be a fine, regular, star triangulation of a reflexive three-dimensional polytope $\Delta^\circ$, and let $\Sigma_1$ be the associated fan. This corresponds to a compact, smooth toric threefold $X_{\Sigma_1}.$ Each minimal generator $u_{\rho}$ for $\rho \in \Sigma(1)$ is said to have ``height 1''. When one blows up $X_{\Sigma_1}$ at a toric curve or point, one obtains a toric variety $X_{\Sigma_2}$ with an exceptional divisor $E$ corresponding to ray $\epsilon \in \Sigma_2(1)$ with minimal generator
\begin{equation}
e = a u_{\rho_1} + b u_{\rho_2} + c u_{\rho_3}\,,
\end{equation} for $a, b, c \in \{0, 1\}$ and $u_{\rho_i} \in \Sigma_1(1)$ for $i = 1, 2, 3.$ We then set the height of $e$ to be 
\begin{align}
\text{height}(e) & = a \times \text{height}(u_{\rho_1}) + b \times \text{height}(u_{\rho_2}) +c \times \text{height}(u_{\rho_3}) \label{eq:height_equation}\\ & = a + b + c\, . 
\end{align} Thus the minimal generator of every ray in $\Sigma_2$ is assigned a height, and then by iteratively blowing up toric curves and points in $X_{\Sigma_2}$ and applying equation \eqref{eq:height_equation}, one obtains a sequence of toric varieties $X_{\Sigma_1}, X_{\Sigma_2}, X_{\Sigma_3}, \ldots$ where if $\rho \in \Sigma_i$ is a ray, then its corresponding minimal generator $u_\rho$ has a well-defined height. 

The most important fact is that if all minimal generators $u_\rho$ of rays  $\rho \in \Sigma_i(1)$ for any $i$ satisfy $\text{height}(u_\rho) \leq 6$, then the polar dual of $\Delta^\circ$ is contained in $\mathcal{P}_{\mathcal{G}_i}$, where $\mathcal{P}_{\mathcal{G}_i}$ is defined in equation \eqref{eq:p_definition}. In other words, by Proposition \ref{prop:poly} for generic choice of $f \in \mathcal{O}_{X_{\Sigma_i}}(-4K_{X_{\Sigma_i}})$ and $g \in \mathcal{O}_{X_{\Sigma_i}}(-6K_{X_{\Sigma_i}})$, $X_{\Sigma_i}$ satisfies equation \eqref{eqn:finite_distance} for all toric subvarieties. To see this, let $u_\varrho$ be an arbitrary minimal generator. Since it has height at most 6, we can write 
\begin{equation}
u_\varrho = \sum_{\rho \in \Sigma_1} a_\rho u_\rho\,,
\end{equation} 
for natural numbers $a_\rho$ satisfying $\sum_{\rho \in \Sigma_1} a_\rho \leq 6$, and for $u_\rho$ the minimal generators of the rays $\rho \in \Sigma_1(1).$ Then let 
\begin{equation}
\Delta = \{m \in M | \langle m, n \rangle \geq -1 \text{ for all }n \in \Delta^\circ\}\,,
\end{equation} 
be the polar dual of $\Delta^\circ$, and let $m \in \Delta$. Then 
\begin{align}
\langle m, u_\varrho \rangle & = \sum_{\rho \in \Sigma} a_\rho \langle m, u_\rho \rangle \\ & \geq -6\,,
\end{align} 
since $\langle m, u_\rho\rangle \geq -1$ for all $\rho.$ However, that means that $m \in \mathcal{G}$, and so all integer points of $\Delta$ are contained in $\mathcal{G}.$ In other words, $\Delta \subseteq \mathcal{P}_{\mathcal{G}_i},$ and so the origin is contained in the strict interior of $\mathcal{P}_{\mathcal{G}_i}.$ Hence, $X_{\Sigma_i}$ satisfies \eqref{eqn:finite_distance}.

The previous paragraph established that any sequence of blowups of $X_{\Sigma_1}$ yields a base satisfying equation \eqref{eqn:finite_distance} on all toric subvarieties, provided that the minimal generator of every prime toric divisor has height at most 6. However, to have some control over what is going on, we do the blowups in a specific order. In particular, we first add so called ``face trees'' to the toric base, and then add ``edge trees.'' 

To describe face trees, suppose $\sigma \in \Sigma_1(3)$ is a maximal cone. Then since $X_{\Sigma_1}$ is smooth, this cone is the collection of points that are non-negative $\mathbb{R}$-linear combinations of three ray minimal generators, say $u_{\rho_1}, u_{\rho_2},$ and $u_{\rho_3}.$ A face tree $F_\sigma$ is a series of toric blowups of the toric point corresponding to $\sigma$ satisfying the following condition: if $E$ is one of the exceptional prime toric divisors, then its corresponding ray minimal generator, say $e$, has height at most 6, and can be written as \begin{equation}e = a u_{\rho_1} + b u_{\rho_2} + c u_{\rho_2}\,,\end{equation} for $a, b, c \in \mathbb{N}$ with $a, b, c \geq 1.$ The condition that $a, b, c \geq 1$ ensures that the face tree blowup does not touch the two-dimensional cones at the boundary of $\sigma.$ One can calculate that there are $41{,}873{,}645$ face trees. 

To describe edge trees, suppose $\tau \in \Sigma(2).$ Then $\tau$ is the collection of points which are non-negative $\mathbb{R}$-linear combinations of two ray minimal generators, say $u_{\rho_1}$ and $u_{\rho_2}.$ An edge tree $E_\tau$ is then a series of toric blowups of the toric curve corresponding to $\tau$ satisfying the following condition: if $E$ is one of the exceptional prime toric divisors, then its corresponding ray minimal generator, say $e$, has height at most 6. A straightforward calculation shows that there are 82 edge trees.

The Tree ensemble is then constructed as follows. Fix a fine, regular, star triangulation of each of the 4,319 three-dimensional reflexive polytopes which give rise to weak Fano toric varieties. Then, for each maximal cone $\sigma \in \Sigma(3)$, and for each two-cone $\tau \in \Sigma(2)$, choose, uniformly at random, a face tree $F_\sigma$ and an edge tree $E_\tau$. Then obtain a new fan $\Sigma'$ by first replacing all $\sigma \in \Sigma(3)$ with their corresponding $F_\sigma,$ and then replace all $\tau \in \Sigma(2)$ with their corresponding $E_\tau.$ This produces a toric variety $X_{\Sigma'}$ in the Tree ensemble. In our work we restrict attention to the sub-ensemble consisting of trees built on either of the two reflexive polytopes yielding $h^{1,1} = 35$ weak-Fano toric varieties.

We direct the reader to the figures in \cite{trees} for a pictorial understanding of the construction.

\subsubsection{Skeleton Ensemble}\label{appsec:skeleton}
In the Skeleton ensemble, one performs a Monte Carlo as follows. Given a smooth compact toric threefold $X_\Sigma$ with fan $\Sigma$, such that $(0, 0, 0)$ is in the strict interior of $\mathcal{P}_{\mathcal{G}}$, as defined in equation \eqref{eq:p_definition}, one then
\begin{enumerate}
    \item Lists all potential toric curve and point blowups of $X_\Sigma$.
    \item Randomly chooses one of them to perform, producing a new toric variety $X_{\Sigma'}$ with associated $\mathcal{P}_{\mathcal{G}_i}$.
    \item  Accepts the new geometry as the next step of the Monte Carlo chain if $(0, 0, 0)$ is in the strict interior of $\mathcal{P}_{\mathcal{G}'}.$ Otherwise, one rejects the geometry, and performs another random blowup.
\end{enumerate}
Since there are only finitely many topological types \cite{DiCerboSvaldi} of elliptic Calabi-Yau fourfolds satisfying equation \eqref{eqn:finite_distance}, this algorithm terminates. 
\begin{table}
\small
\begin{center}
\hspace*{-1.5cm}
\begin{tabular}{ |c|c||c|c||c|c||c|c||c|c||c|c| } 
 \hline
 $h^{1,1}(B_3)$ & $n_\text{final}$ & 
 $h^{1,1}(B_3)$ & $n_\text{final}$ &
 $h^{1,1}(B_3)$ & $n_\text{final}$ & 
 $h^{1,1}(B_3)$ & $n_\text{final}$ &
 $h^{1,1}(B_3)$ & $n_\text{final}$ &
 $h^{1,1}(B_3)$ & $n_\text{final}$ \\
 \hline

999\;\; & 2 & 
3276 $\star$ & 112 & 
4463\;\;\; & 6 & 
5674\;\;\; & 2 & 
7401\;\;\; & 7 & 
10170\;\;\; & 1 \\
1151\;\;\; & 3 & 
3284\;\;\; & 3 & 
4468 $\star$ & 141 & 
5728\;\;\; & 1 & 
7425\;\;\; & 4 & 
10421\;\;\; & 1 \\
1596 \;\; & 6 & 
3295 $\star$ & 323 & 
4470\;\;\; & 4 & 
5851\;\;\; & 2 & 
7478\;\;\; & 1 & 
10480\;\;\; & 9 \\
1727 $\star$ & 316 & 
3332\;\;\; & 18 & 
4520 $\star$ & 4 & 
5855\;\;\; & 18 & 
7498 $\star$ & 101 & 
10573\;\;\; & 3 \\
1799 $\star$ & 31 & 
3374 $\star$ & 101 & 
4563\;\;\; & 10 & 
5859\;\;\; & 8 & 
7516\;\;\; & 27 & 
10618\;\;\; & 2 \\
1882 $\star$ & 8 & 
3401 $\star$ & 139 & 
4602\;\;\; & 2 & 
5878 $\star$ & 33 & 
7526 $\star$ & 2 & 
10646\;\;\; & 13 \\
1943 $\star$ & 4613 & 
3422 $\star$ & 793 & 
4642\;\;\; & 2 & 
5880\;\;\; & 2 & 
7658\;\;\; & 1 & 
10723\;\;\; & 3 \\
2015 $\star$ & 722 & 
3427\;\;\; & 5 & 
4685\;\;\; & 3 & 
5936\;\;\; & 1 & 
7692\;\;\; & 1 & 
10734\;\;\; & 2 \\
2047 $\star$ & 668 & 
3498 $\star$ & 52 & 
4687\;\;\; & 12 & 
5942\;\;\; & 4 & 
7700\;\;\; & 1 & 
10974\;\;\; & 1 \\
2057 $\star$ & 3575 & 
3527\;\;\; & 29 & 
4703\;\;\; & 1 & 
5989 $\star$ & 311 & 
7771\;\;\; & 2 & 
11341 $\star$ & 34 \\
2119\;\;\; & 2 & 
3539 $\star$ & 78 & 
4724\;\;\; & 4 & 
5998\;\;\; & 1 & 
7774\;\;\; & 1 & 
11451\;\;\; & 3 \\
2186 $\star$ & 1043 & 
3558\;\;\; & 1 & 
4741 $\star$ & 7 & 
6070\;\;\; & 16 & 
7792\;\;\; & 1 & 
11506\;\;\; & 1 \\
2199 $\star$ & 197 & 
3583\;\;\; & 2 & 
4748 $\star$ & 42 & 
6125\;\;\; & 1 & 
7909 $\star$ & 136 & 
11524\;\;\; & 8 \\
2249 $\star$ & 7658 & 
3599 $\star$ & 5 & 
4762\;\;\; & 12 & 
6143 $\star$ & 64 & 
7911\;\;\; & 1 & 
11583\;\;\; & 3 \\
2303 $\star$ & 7563 & 
3658 $\star$ & 243 & 
4863\;\;\; & 13 & 
6152\;\;\; & 1 & 
7981\;\;\; & 3 & 
11717\;\;\; & 1 \\
2395 $\star$ & 357 & 
3686 $\star$ & 1738 & 
4866\;\;\; & 7 & 
6216\;\;\; & 4 & 
8111 $\star$ & 23 & 
11977\;\;\; & 2 \\
2399 $\star$ & 951 & 
3701\;\;\; & 3 & 
4913 $\star$ & 102 & 
6241\;\;\; & 1 & 
8114\;\;\; & 1 & 
12157\;\;\; & 2 \\
2429\;\;\; & 22 & 
3739 $\star$ & 25 & 
4920\;\;\; & 6 & 
6250\;\;\; & 9 & 
8132\;\;\; & 1 & 
12192\;\;\; & 4 \\
2491 $\star$ & 58 & 
3741 $\star$ & 34 & 
4938\;\;\; & 1 & 
6271\;\;\; & 2 & 
8191\;\;\; & 1 & 
12366\;\;\; & 1 \\
2591 $\star$ & 4777 & 
3789 $\star$ & 13 & 
4939 $\star$ & 271 & 
6346\;\;\; & 1 & 
8212\;\;\; & 6 & 
12439\;\;\; & 2 \\
2599 $\star$ & 373 & 
3811 $\star$ & 58 & 
4946 $\star$ & 5 & 
6384\;\;\; & 1 & 
8230 $\star$ & 58 & 
12530\;\;\; & 1 \\
2623 $\star$ & 1662 & 
3817 $\star$ & 30 & 
4988\;\;\; & 2 & 
6399\;\;\; & 25 & 
8435 $\star$ & 4 & 
12631 $\star$ & 16 \\
2636 $\star$ & 108 & 
3872\;\;\; & 5 & 
5063\;\;\; & 2 & 
6414\;\;\; & 1 & 
8452\;\;\; & 3 & 
12743\;\;\; & 1 \\
2661 $\star$ & 693 & 
3887 $\star$ & 454 & 
5113\;\;\; & 31 & 
6440 $\star$ & 10 & 
8676\;\;\; & 2 & 
13073\;\;\; & 1 \\
2679\;\;\; & 2 & 
3968\;\;\; & 27 & 
5143 $\star$ & 608 & 
6586\;\;\; & 2 & 
8730\;\;\; & 22 & 
13344\;\;\; & 1 \\
2683\;\;\; & 9 & 
3992 $\star$ & 12 & 
5165\;\;\; & 7 & 
6614\;\;\; & 3 & 
8926\;\;\; & 2 & 
13372\;\;\; & 1 \\
2743\;\;\; & 16 & 
3999\;\;\; & 16 & 
5183\;\;\; & 4 & 
6638\;\;\; & 2 & 
8938 $\star$ & 2 & 
13390\;\;\; & 3 \\
2821 $\star$ & 123 & 
4019\;\;\; & 78 & 
5211\;\;\; & 1 & 
6660\;\;\; & 2 & 
8980 $\star$ & 126 & 
13529\;\;\; & 1 \\
2824 $\star$ & 343 & 
4049 $\star$ & 149 & 
5316\;\;\; & 3 & 
6784 $\star$ & 8 & 
8999 $\star$ & 41 & 
13628\;\;\; & 1 \\
2891 $\star$ & 26 & 
4056\;\;\; & 6 & 
5356 $\star$ & 5 & 
6802 $\star$ & 72 & 
9374\;\;\; & 18 & 
13994\;\;\; & 6 \\
2915 $\star$ & 73 & 
4115\;\;\; & 26 & 
5383 $\star$ & 74 & 
6857\;\;\; & 1 & 
9439\;\;\; & 2 & 
14378\;\;\; & 3 \\
2943 $\star$ & 1 & 
4147\;\;\; & 1 & 
5416\;\;\; & 6 & 
6892\;\;\; & 1 & 
9528\;\;\; & 2 & 
14561\;\;\; & 1 \\
2961 $\star$ & 440 & 
4211 $\star$ & 27 & 
5422\;\;\; & 1 & 
6911 $\star$ & 131 & 
9543\;\;\; & 6 & 
14774\;\;\; & 1 \\
2999 $\star$ & 1218 & 
4223\;\;\; & 3 & 
5460\;\;\; & 1 & 
6945 $\star$ & 18 & 
9640\;\;\; & 1 & 
15402\;\;\; & 1 \\
3010\;\;\; & 5 & 
4253\;\;\; & 4 & 
5476\;\;\; & 17 & 
6968\;\;\; & 3 & 
9786\;\;\; & 4 & 
15430\;\;\; & 4 \\
3037 $\star$ & 163 & 
4274 $\star$ & 12 & 
5487\;\;\; & 2 & 
7031\;\;\; & 1 & 
9804\;\;\; & 26 & 
15970\;\;\; & 1 \\
3071 $\star$ & 57 & 
4373 $\star$ & 796 & 
5503 $\star$ & 25 & 
7087\;\;\; & 4 & 
9952\;\;\; & 4 & 
- & - \\
3086 $\star$ & 2774 & 
4375 $\star$ & 82 & 
5515\;\;\; & 10 & 
7206\;\;\; & 2 & 
9998\;\;\; & 2 & 
- & - \\
3157 $\star$ & 11 & 
4394 $\star$ & 385 & 
5522 $\star$ & 163 & 
7328\;\;\; & 1 & 
10042\;\;\; & 17 & 
- & - \\
3247 $\star$ & 19 & 
4430\;\;\; & 27 & 
5557\;\;\; & 37 & 
7363\;\;\; & 1 & 
10081\;\;\; & 1 & 
- & - \\
3275\;\;\; & 80 & 
4443\;\;\; & 10 & 
5644\;\;\; & 2 & 
7373 $\star$ & 4 & 
10124 $\star$ & 73 & 
- & - \\
 
 \hline
\end{tabular}
\caption{$h^{1,1}$ of final bases obtained in 50,000 Monte-Carlo runs starting from $\mathbb{P}^3$. Starred $h^{1,1}$ are those final base $h^{1,1}$ that are represented in table 2 of \cite{skeleton}. The only final base $h^{1,1}$ values that appear in \cite{skeleton} but not here are $h^{1,1} = 1{,}317$ and $h^{1,1} = 5{,}623$.}
\end{center}
\end{table}
\subsubsection{K2 Ensemble}\label{sec:k2ens}
Reference \cite{Yinan} provides a fan $\Sigma_{B_3}$ yielding a smooth, compact toric threefold $B_3$. $B_3$ satisfies the hypotheses of proposition \ref{prop:poly}, and so that $B_3$, satisfies condition \eqref{eqn:finite_distance} on all toric subvarieties. Moreover, this means that any toric threefold obtained from a fan whose rays are a subset of $\Sigma(1)$ also satisfies \eqref{eqn:finite_distance}. Hence, we can construct an ensemble of smooth, compact toric threefold bases satisfying condition \eqref{eqn:finite_distance} on all toric subvarieties by first placing face trees on all 5,016 toric (8,12) points of $B_\text{seed},$ and then placing edge trees on all 7{,}576 toric (4,6) curves of $B_\text{seed}.$ We also include $B_3$ from \cite{Yinan} \S2 in the ensemble. Since there are 19 different edge trees and 41,873,645 different face trees, this yields an ensemble of \begin{align}\# \text{ K2 Ensemble} & = 19^{7576} \times 41,873,645^{5{,}016} + 1 \\ & = 5.11 \times 10^{52{,}730}\,,\end{align} elements.\footnote{This assumes there are no nontrivial $\text{GL}_N(\mathbb{Z})$ fan automorphisms.} Due to the combinatorial nature of the ensemble, one can sample uniformly at random from it.

\subsection{Sampling Method}
In this section we outline how we obtained the points in Kähler moduli space we used to evaluate axion masses, decay constants, and effective couplings to photons. Broadly, for each geometry in the KS, Tree, and Skeleton ensembles, we: 
\begin{enumerate}
    \item Computed the tip of the dual Coxeter stretched cone defined in \eqref{eq:dcskc_def}, and computed axion masses and decay constants there.
    \item Sampled 20 pairings of $\bigl(D_\text{EM},~\text{vol}(D_\text{EM})\bigr)$.
    \item For each choice of $\bigl(D_\text{EM},~\text{vol}(D_\text{EM})\bigr)$, computed the $\ell^p$ minimizing point of the subregion of the dual Coxeter stretched K\"ahler cone where $D_\text{EM}$ took on the chosen volume, and computed the effective axion-photon coupling there for comparison to helioscope and X-ray experiments. 
\end{enumerate}

In the KS ensemble with $h^{1,1} \leq 174$, the Tree ensemble, and the Skeleton ensemble with $h^{1,1} \leq 2{,}207$ we compute the $\ell^2$ tips of the above cones. Otherwise, due to the computational complexity, we work at the $\ell^1$ tips. 
 
To accomplish (1) in the case of a toric threefold base, $B_3 \in S$ with fan $\Sigma$, and ray minimal generators $u_\rho$ for $\rho \in \Sigma(1),$ we compute $\mathcal{F}, \mathcal{G}, f \in \mathcal{O}_{B_3}(-4K_{B_3}),$ and $g\in \mathcal{O}_{B_3}(-6K_{B_3})$ as in equations \eqref{eq:f_poly_def} through \eqref{eq:g_section_real_def}, taking $f, g$ to be generic.   

On each prime toric divisor $D_\rho$ for $\rho \in \Sigma(1)$, we then compute the generic order of vanishing of $f$ as 
\begin{equation}\text{ord}_{D_\rho}(f) = \min_{m \in \mathcal{F}} \{\langle m, u_\rho \rangle + 4\}\,,
\end{equation} and the generic order of vanishing of $g$ as \begin{equation}\text{ord}_{D_\rho}(g) = \min_{m \in \mathcal{G}} \{\langle m, u_\rho \rangle + 6\}\,.
\end{equation} This means that the order of vanishing of the discriminant polynomial 
\begin{equation}\Delta = 4 f^3 + 27 g^2\,,
\end{equation} is given by 
\begin{equation}\text{ord}_{D_\rho}(\Delta) = \min \bigg(3 \times \text{ord}_{D_\rho}(f), 2 \times \text{ord}_{D_\rho}(g)\bigg)\,.
\end{equation}Comparison with Table \ref{tab:gauge_groups}  
\begin{table}
    \centering
    \begin{tabular}{ c c c c c c }
     fiber type & ord(f) & ord(g) & ord($\Delta$) & sing. & $G$ \\ \hline II & $\geq$ 1 & 1 & 2 & - &  - \\ III & 1 & $\geq$ 2 & 3 & $A_1$ & SU(2) \\ IV & $\geq$ 2  & 2 & 4 & $A_2$  & Sp(1) or SU(3) \\ $I_0^*$ & $\geq$ 2 & $\geq 3$ & 6 & $D_4$ & $G_2$ or SO(7) or SO(8) \\ $IV^*$ & $\geq 3$ & 4 & 8 & $E_6$ & $F_4$ or $E_6$ \\ $III^*$ & 3 & $\geq $ 5 & 9 & $E_7$ & $E_7$ \\ $II^*$ & $\geq 4$ & 5 & 10 & $E_8$ & $E_8$ 
    \end{tabular}
    \caption{Table of possible non-Higgsable gauge groups. The ``fiber type'' refers to the Kodaira fiber type; ord(f), ord(g), and ord($\Delta$) are the vanishing orders of generic $f \in \mathcal{O}(-4K)$, $g \in \mathcal{O}(-6K)$, and $\Delta = 4f^3 + 27 g^2$; ``sing.'' is the type of fibral singularity; and $G$ is the gauge group. Where multiple gauge groups are listed, the gauge group can vary depending on whether the fiber is split or not: see \cite{Bershadsky:1996nh,Grassi:2018wfy}.}
    \label{tab:gauge_groups}
    \end{table}
allows one to read off the gauge group on $D_\rho$, which we denote $G_\rho$,\footnote{Strictly speaking, one could tune the coefficients $a_m$ and $b_m$ to enhance the vanishing orders of $f, g,$ or $\Delta$ on particular prime toric divisors, and obtain other gauge groups. However, in our analysis, we do not perform such tunings.} and from the gauge group one then reads off the associated dual Coxeter number from Table \ref{tab:dcs}. 
\begin{table}
    \centering
    \begin{tabular}{ c | c c c c c c c c c c}
     & $\text{SU}(2)$ & Sp(1) & SU(3) & $G_2$ & SO(7) & SO(8) & $F_4$ & $E_6$ & $E_7$ & $E_8$ \\\hline $c_2(G)$ & 2 & 2 & 3 & 4 & 6 & 6 & 9 & 12 & 18 & 30 \\ $\text{dim}_{\mathbb{R}}(G)$ & 3 & 3 & 8 & 28 & 21 & 56 & 104 & 156 & 266 & 496 
    \end{tabular}
    \caption{Dual Coxeter numbers $c_2(G)$ and real dimensions $\text{dim}_{\mathbb{R}}(G)$ of non-Higgsable gauge groups $G$.  The real dimension corresponds to the number of gauge bosons.}
    \label{tab:dcs}
    \end{table}
We then compute the K\"ahler cone of $B_3$, which we call $\mathcal{K}$.  If cycles in $B_3$ have volume comparable to the string scale, then instantons and other stringy effects become appreciable. 
Hence, to have control of the effective field theory, we restrict attention to the ``dual Coxeter stretched K\"ahler cone'', denoted $\mathcal{K}_{c_2}$, and defined to be the collection of K\"ahler forms $J \in \mathcal{K}$ with 
\begin{equation}\label{eq:dcskc_def}
\text{vol}(C) = \int_{C} J \geq 1\qquad \text{and} \qquad\text{vol}(D_\rho) = \frac{1}{2} \int_{[D_\rho]} J \wedge J \geq c_2(G_\rho)\,, \end{equation} 
for all holomorphic curves $C$, and   
for all $\rho \in \Sigma(1)$. 
Here $c_2(G_\rho)$ is the dual Coxeter number of the non-Higgsable gauge group on $D_{\rho}$, if one is present, and $c_2(G_\rho) = 1$
otherwise.

As we prove in Appendix 
\ref{sec:convex}, the set $\mathcal{K}_{c_2}$ obeying \eqref{eq:dcskc_def} is convex.\footnote{In the KS ensemble, we explicitly verified that the intersection numbers of every prime toric divisor of every Calabi-Yau in our ensemble satisified the conclusions of Lemma \ref{lemma:eig}.} If we let $N = h^{1,1}(B_3)$; order the prime torics, say $D_1, \ldots, D_{N+3}$; and let $\omega^1, \ldots, \omega^{N+3} \in H^{1,1}(B_3, \mathbb{Z})$ be the (1, 1)-forms Poincar\'e dual to $[D_\rho] \in H_4(B_3, \mathbb{Z})$; then one can expand any K\"ahler form $J \in \mathcal{K}$, and in particular any $J \in \mathcal{K}_{c_2}$ as 
\begin{equation}
J = \sum_{a = 1}^{N+3} t_a \omega^a\, .
\label{eq:dcskc_expansion}
\end{equation} 
One then defines the tip of $\mathcal{K}_{c_2}$ to be the $J_\star \in \mathcal{K}_{c_2}$ whose parameters $t_a$ solve the quadratic program 
\begin{align}
\min \sum_{a = 1}^{N+3} t_a^2\, , \\ \text{s.t. } \sum_{c = 1}^{N+3} \kappa^{abc} t_c \geq 1 & \; \forall 1 \leq a < b \leq N + 3 \text{ with }D_a \cap D_b \neq \varnothing\,, \\ \frac{1}{2}\sum_{b, c = 1}^{N+3} \kappa^{abc} t_b t_c \geq c_2(D_a) & \; \forall 1 \leq a \leq N + 3\,, \\ t_a \in \mathbb{R} & \; \forall 1 \leq a \leq N + 3\,.
\end{align} 
Note that even though the expansion in equation \eqref{eq:dcskc_expansion} is not unique for a given K\"ahler form $J$, the parameters $t_a$ determined from the quadratic program will be unique.\footnote{One could find a $\mathbb{Z}$-basis for the homology and then do the above procedure in-basis, but then the tip would also depend on the basis choice.}

We then find all pairings $(D_{\rho_0}, \tau_\text{EM})$ where $D_{\rho_0}$ is a prime toric divisor of the toric variety hosting a non-higgsable gauge group which we denote $G_{\rho_0}$, and $\tau_\text{EM} \in \{10, 20, 30, 40, 50\}$ is greater than the volume of $D_\rho$ at $J_\star.$ For each geometry we then choose 20 such pairings uniformly at random. For each choice, we assume $G_{\rho_0}$ is broken to $U(1)_\text{EM}$ at high scale, and set $c_2(G_{\rho_0}) = 1$. We then compute the tip of the region in the dual-coxeter stretched cone where $\text{vol}(D_{\rho_0})$ achieves volume $\tau_{\text{EM}}$ as in appendix \ref{sec:candc}. Enforcing that the volume of the divisor hosting $U(1)_\text{EM}$ is $\tau_{\text{EM}}$ enforces that the UV coupling \begin{equation}
\alpha_\text{UV} = \frac{1}{\text{vol}(D_{\rho_0})}\,,
\end{equation} 
lies in the desired range 
\begin{equation}
\alpha_\text{UV}^{-1} \in \{10,20,30,40,50\}\,.
\end{equation}

\section{Canonical Couplings}
\label{app:couplings}

In this Appendix we explain in detail how the axion Lagrangian is obtained from topological and geometric data.
 
We begin by standardizing notation. We compactify F-theory on a smooth K\"ahler threefold with $N = h^{1,1}(B_3)$. Let $X_1, \ldots, X_N$ be a basis 
for $H_4(B_3, \mathbb{Z})$,
with Poincar\'e-dual $(1,1)$-forms $\omega^1, \ldots, \omega^N \in H^{1,1}(B_3, \mathbb{Z}).$ Then the triple-intersection numbers are given by \begin{equation}\label{eq:tripleint}
\kappa^{abc} := \int_{B_3} \omega^a \wedge \omega^b \wedge \omega^c\,,
\end{equation} where $a,b,c$ range from $1$ to $N$, and one can expand the K\"ahler form $J$ as 
\begin{equation}J = \sum_{a = 1}^N t_a \omega^a\,,\end{equation} where $t_a$ are the K\"ahler parameters. The overall Einstein frame  
volume is 
\begin{align}\mathcal{V}  := \frac{1}{6}\int_{B_3} J \wedge J \wedge J = \frac{1}{6}\sum_{a,b,c = 1}^N \kappa^{abc} t_a t_b t_c\,,
\end{align} 
and the volume of a holomorphic 4-cycle $X_a$ is 
\begin{align}\tau^a   := \frac{1}{2} \int_{X_a} J \wedge J   = \frac{1}{2} \int_{B_3} J \wedge J \wedge \omega^a  = \frac{1}{2} \sum_{b,c = 1}^N \kappa^{abc} t_b t_c\,.
\end{align} 
\subsection{Kinetic Terms} 

The bosonic part of the supergravity action is 
\begin{align}S \supset \int d^4x \sqrt{-g} \bigg(\frac{1}{2}R - K_{a\bar{b}}\,\partial_\mu \phi^a \partial^\mu \bar{\phi}^{\bar{b}} - V_F\bigg)\,,\end{align} where $\phi^a$ are complex scalar fields, 
\begin{align}
K_{a\bar{b}} & = \frac{\partial}{\partial \phi^a} \frac{\partial}{\partial \bar{\phi}^{\bar{b}}} \mathcal{K}\,,
\end{align} 
is the K\"ahler metric, 
\begin{align}
\mathcal{K} &= -2 \log \mathcal{V} + \mathcal{K}_\text{other}\,,
\end{align} 
is the K\"ahler potential, and 
\begin{align}
V_F & = e^{\mathcal{K}} \left(K^{a\bar{b}}D_a W D_{\bar{b}}W-3|W|^2\right)\,,
\end{align} 
is the F-term potential, where $W$ is the superpotential and \begin{align}
D_a W & = \frac{\partial}{\partial \phi^a} W + \left(\frac{\partial}{\partial \phi^a} \mathcal{K}\right)W\,,
\end{align} 
is the K\"ahler-covariant derivative.

The axion fields $\theta^a$ are buried inside the K\"ahler coordinates $T^a$, which take the form \begin{align}T^a & = \tau^a + i \theta^a\,,\end{align} where $\tau^a$ are the saxion fields.   
Setting $K_{a\bar{b}} = \frac{1}{2} K_{ab}^{(T)}$, so that the kinetic term has the standard factor of $\frac{1}{2}$ out front, the part of the K\"ahler metric corresponding to the K\"ahler parameters is 
\begin{align}
K_{ab}^{(T)}  = \frac{1}{2} \frac{\partial}{\partial \tau^a} \frac{\partial}{\partial \tau^b} \mathcal{K}  = - \frac{\partial}{\partial \tau^a} \left(A_{bc} \frac{1}{\mathcal{V}} \tau^c\right)  = -\frac{1}{2 \mathcal{V}}\left(A_{ab} - \frac{1}{2\mathcal{V}}t_a t_b\right)\,,
\end{align} 
where at intermediate steps we have defined (see e.g.~\cite{Long_2014})
\begin{align}
A_{bc} & := \frac{\partial t_c}{\partial \tau^b}\,,
\end{align} 
where the $t_c$ are the K\"ahler parameters, and we used the identity that 
\begin{align}
A_{ab} \tau^b & = \frac{1}{2} t_a\,.
\end{align} 
The issue is that $A_{bc}$ is not easily written in terms of the K\"ahler parameters. 
However, its inverse is given by 
\begin{align}(A^{-1})^{ab}  = \frac{\partial \tau^b}{\partial t_a}  = \kappa^{abc}t_c\,,
\end{align} 
where $\kappa^{abc}$ are the triple intersection numbers defined in \eqref{eq:tripleint}, and so one verifies that the inverse K\"ahler metric is given by \begin{align}
(K^{-1}_{(T)})^{ab} & = 2 \left(\tau^a \tau^b - \mathcal{V}\kappa^{abc} t_c\right)\,.
\end{align} 

\subsection{Nonperturbative Superpotential}\label{sec:wnp}

We begin with an arbitrary $\mathbb{Q}$-basis $[X_a]$, $a=1,\ldots,N$, for $H_4(B_3, \mathbb{Z})$.
Then a general effective divisor class $[D_A]$ can be written
$[D_A] = \sum_{a} Q_{Aa}[X_a]$, 
with $Q_{Aa} \in \mathbb{Q}$ and with $A \in \mathbb{N}$.
We use $A \in \mathbb{N}$  to formally index all effective divisors;
$a=1,\ldots,N$ to index the $N$ basis divisors;
$I=1,\ldots,N+3$ to index the $N+3$ prime toric divisors;
and $\alpha \subsetneq \{1,\ldots,N\}$ to index the subset of $N_{\text{gauge}}<N$ prime toric divisors that host non-Higgsable gauge groups.

We do not perform a counting of fermion zero modes in this work.  
However, we can parameterize our ignorance by writing the nonperturbative superpotential as a formal sum over \emph{all} effective divisors $D_A$, 
\begin{equation}
\label{eq:wfullapp}
W_{\text{np}}  = \sum_{D_A} \mathcal{A}_A\,  e^{-2 \pi Q_{Ab}T^b/C_2(D_A)}\,,
\end{equation} 
with the understanding that for those $D_A$ with an excess of fermion zero modes, the Pfaffian $\mathcal{A}_A$ should be taken to be zero.  The corresponding terms can be omitted from \eqref{eq:wfullapp},
leading to
\begin{equation}
\label{eq:wfullapp2}
W_{\text{np}}  = \sum_{D_A|\mathcal{A}_A\neq 0}  \mathcal{A}_A\,e^{-2 \pi Q_{Ab}T^b/C_2(D_A)}\,.
\end{equation}

\subsubsection{Prime Toric Model}

At present it is not computationally feasible to enumerate \emph{all} $D_A$ for which $\mathcal{A}_A \neq 0$, so we now lay out a simpler model that approximates \eqref{eq:wfullapp2}.

At a given point in  K\"ahler moduli space, the relative importance of terms in \eqref{eq:wfullapp2}
is controlled --- up to logarithmic corrections from the Pfaffians $\mathcal{A}_A$ --- by
\begin{equation}\label{eq:sdef}
S_A :=2\pi\frac{Q_{Aa} \tau^a}{c_2(D_A)}\,,
\end{equation} with the terms of smallest $S_A$ being dominant. 
One very often finds that the $N$ smallest divisors are prime toric divisors $D_I$.\footnote{Moreover, the dominant set of $N$ contributions produced by the algorithms of \S\ref{app:basis} \emph{always} consists of $N$ of the $D_I$, because every effective divisor can be written as a positive integer linear combination of prime toric divisors.}  
Moreover, one can determine whether
$\mathcal{A}_I \neq 0$ for the $N+3$
prime toric divisors $D_I$, and typically $\mathcal{A}_I \neq 0$ for most of the $D_I$ \cite{Halverson:2019vmd}.

In such a case we have
\begin{equation}
\label{eq:wfullapp3}
W_{\text{np}} \approx \sum_{D_I|\mathcal{A}_I\neq 0} e^{-2 \pi Q_{Ib}T^b/C_2(D_I)}  \approx \sum_{D_I} e^{-2 \pi Q_{Ib}T^b/C_2(D_I)} \,,
\end{equation} where the 
first approximation omits terms from divisors that are effective but not prime toric, and the second approximation is that all (or nearly all) of the prime toric divisors have $\mathcal{A}_I \neq 0$, and for these divisors we take $\mathcal{A}_I=1$.
Throughout this paper, we make the approximation \eqref{eq:wfullapp3}, which we refer to as the \emph{prime toric model}.  
This model should capture all gaugino condensate terms from NHCs, because 
we expect that in our toric setting, 
every NHC occurs on some prime toric divisor $D_\alpha$.

The prime toric model is an approximation that can miss in either direction: a prime toric divisor can be non-rigid, and so not contribute to the superpotential, and on the other hand a nontrivial sum of prime toric divisors can host an important Euclidean D3-brane effect.
Nonetheless, in studies to date, the prime toric model provides a rather accurate approximation at large $h^{1,1}$: see e.g.~\cite{small_ccs}.

A final approximation that is both accurate and convenient 
is to restrict to the $N$ leading contributions in \eqref{eq:wfullapp}, rather than keeping $N+3$ terms.
Then, with a suitable choice of homology basis, the charge matrix $Q_{ab}$ becomes the identity, and the nonperturbative superpotential takes the form 
\begin{equation}
\label{eq:wfullapp4} 
W_{\text{np}}  = \sum_{a=1}^N e^{-2 \pi T^a/C_2(D_a)}\,.
\end{equation} 
In \S\ref{app:basis} we present the algorithm used to perform this truncation and basis choice.

Before proceeding, we remark on the relative importance of Euclidean D3-brane terms in the superpotential versus
the K\"ahler potential.
The superpotential terms result from Euclidean D3-branes wrapping holomorphic four-cycles (effective divisors), whereas K\"ahler potential contributions can result from four-cycles in homology classes that admit no calibrated representatives.
In \cite{Demirtas:2019lfi} the notion of \emph{recombination} was introduced: 
for a homology class $[\Sigma] \in H_4(B_3,\mathbb{R})$
represented by a union $\Sigma_{\cup}$ of holomorphic and antiholomorphic cycles, and with $\Sigma_{\mathrm{min}}$ the minimum-volume representative of $[\Sigma]$, the recombination factor is defined as
\begin{equation}
    \mathfrak{r}:= \frac{\mathrm{Vol}(\Sigma_{\cup})-\mathrm{Vol}(\Sigma_{\mathrm{min}})}{\mathrm{Vol}(\Sigma_{\mathrm{min}})}\,.
\end{equation}
In the absence of significant recombination, i.e.~for $\mathfrak{r}\ll 1$, calibrated volumes $\mathrm{Vol}(\Sigma_{\cup})$ provide a useful approximation to the actual minimal volumes 
$\mathrm{Vol}(\Sigma_{\mathrm{min}})$,
and in such cases K\"ahler potential terms are typically subleading in comparison to superpotential terms.  This conclusion is strengthened when the scale of supersymmetry breaking, as measured by the classical flux superpotential, is relatively small: see the discussion in \cite{Demirtas:2021gsq}.
Because studies to date have
found limited evidence for recombination \cite{Demirtas:2019lfi,Long:2021lon}, in this paper we will not incorporate nonperturbative corrections to the K\"ahler potential.

\subsubsection{Determining the Basis}\label{app:basis}

Suppose that $ \check{D}_1, \ldots,  \check{D}_N$ is an initial set of $N$ prime toric divisors whose classes furnish a 
$\mathbb{Q}$-basis for $H_4(B_3, \mathbb{Z})$, and that $\check{D}_{N+1}$, $\check{D}_{N+2}$,
and $\check{D}_{N+3}$ are the three additional prime toric divisors.\footnote{The analysis presented here can be adapted to our ensemble of Calabi-Yau threefold hypersurfaces in toric fourfolds, with all instances of $+3$ replaced by $+4$.} 
 Under our assumptions, each of $\check{D}_1,\ldots,\check{D}_{N+3}$ supports a superpotential contribution.

Given a point $\vec{t}$ in K\"ahler moduli space, we would like to compute a set $\{D_a\}=\{D_1, \ldots, D_N\}$, with the properties
\begin{enumerate}
    \item Each of the $D_a$ is a prime toric divisor.
    \item $[D_1], \ldots, [D_N]$ furnish a $\mathbb{Q}$-basis for $H_4(B_3, \mathbb{Z})$.
    \item At $\vec{t}$, the superpotential terms from the $D_a$ are the $N$ most important such terms in the axion effective theory. 
\end{enumerate} 
To accomplish this, we write the $N+3$ prime toric divisors in terms of the initial basis as
\begin{equation}
[D_I] = \sum_{a = 1}^N Q_{Ia} [\check{D}_a]\,,\qquad I=1,\ldots,N+3\,.\end{equation}
In terms of \eqref{eq:sdef},
the
the contribution of the instanton wrapping $D_I$ to the F-term potential can be written as   
\begin{equation}
V_F \supset \Lambda_I^4 \cos\Bigl(2 \pi Q_{Ia} \theta^a + \delta_I\Bigr)\,,
\end{equation} 
where 
\begin{equation}
\Lambda_I^4 = \frac{g_s^4}{32} \times \frac{W_0 A_I}{\mathcal{V}^2} S_I\,e^{-S_I}\,,
\end{equation} 
where $g_s$ is the string coupling; $A_I$ is the modulus of the Pfaffian; $W_0$ is the flux part of the superpotential; $\mathcal{V}$ is the overall volume of $X_3$, and $\delta_I$ is the phase of the Pfaffian. We set $W_0 = A_I = 1$ henceforth.

Next, we reorder to sort the instanton scales $\Lambda_{I}$,
such that $\Lambda_{1}^4 \geq \Lambda_{2}^4 \geq \ldots \geq \Lambda_{N+3}^4$.
At this point we can explain what is meant by ``the $N$ most important such terms'' in Condition (3) above.
Naively, one might expect to truncate to the terms involving $\{\Lambda_1,\ldots, \Lambda_N\}$.
However, suppose for example that $\vec{Q}_3$, the vector with components $Q_{3a}$, is in the linear span of 
$\vec{Q}_1$ and
$\vec{Q}_2$.  Then the axion potential term proportional to $\Lambda_3^4$ does not lift a flat direction: it merely yields a correction to the more-leading terms involving $\Lambda_1^4$ and $\Lambda_2^4$.  To collect the $N$ terms furnishing a basis of leading effects, one should ignore $\vec{Q}_3$ and proceed to check $\vec{Q}_4$.\footnote{See e.g.~\cite{Demirtas:2021gsq} for a discussion of related considerations.}

Thus, we propose the following algorithm  
for finding the $N$ instantons $\Lambda^4_I$ yielding the instanton basis.
\begin{center}
\begin{algorithm}[H]
    \DontPrintSemicolon 
    $S \gets \{I_1\}$\;
    $j \gets 2$\;
    \While{$j \leq N+3$}{
        \uIf{$Q_{I_j} \not \in \text{Span}\{Q_{I_i} | i \in S\}$}{
            $S \gets S \cup \{I_j\}$
        }
        $j \gets j + 1$\;   
    }
    \Return S
    \caption{Simple Instanton Basis Determination}
    \label{alg:old_instanton_basis_determination}
\end{algorithm}
\end{center}

Algorithm \ref{alg:old_instanton_basis_determination} is theoretically appealing because one goes from the largest to the smallest instanton from a prime toric divisor, and checks whether each one is subleading or not. However, this is inefficient in practice. Checking the span at step $j$ requires computing the rank of a $j \times N$ matrix, and this must be done $N + 3$ times.

Instead, we use the following algorithm. To standardize notation, let $X_3$ have fan $\Sigma$, with rays $\rho_1, \ldots, \rho_{N+3} \in \Sigma(1)$ corresponding to the prime toric divisors $D_{i_1}, \ldots, D_{i_{N+3}}$,\footnote{The $D_{i_1}, \ldots, D_{i_{N+3}}$ are in general a permutation of $D_1, \ldots, D_{N+3}$.} such that the instanton wrapping $D_{i_j}$ has instanton scale $\Lambda_{I_j}^4$. We let the $u_{\rho_i}$ be the minimal generators of the rays $\rho_i$, and we will 1-index matrices. 
\begin{center}
\begin{algorithm}[H]
    \DontPrintSemicolon
    $M \gets $ the $3 \times (N + 3)$ matrix whose $i$th column is $u_{\rho_{N + 3 - (i- 1)}}.$ \; 
    $Q, R \gets $ QR decomposition of $M$ \;
    $S' \gets \varnothing$\;
    $j \gets 1$\;
    \While{$j \leq 3$}{
        $i \gets $ index of first nonzero column in row $j$ of $R$ \;
        $S' \gets S' \cup \{I_{N + 3 - (i-1)}\}$ \;
        $j \gets j + 1$\;
    }
    $S \gets \{I_1, \ldots, I_{N + 3}\} \setminus S'$\;
    \Return $S$
    \caption{Fast Instanton Basis Determination}
    \label{alg:new_instanton_basis_determination}
\end{algorithm}
\end{center}

Here, QR decomposition returns a $3 \times 3$ orthogonal matrix $Q$ and a $3 \times (N + 3)$ upper-triangular matrix $R$ such that 
\begin{equation}M = QR\,.\end{equation} The rows of $M$ give the linear relations among the divisor classes $[D_{i_a}]$ in the sense that \begin{equation}\sum_{a = 1}^{N+3} M_{ja} [D_{i_{N + 3 - (a-1)}}] = 0\,,\end{equation} for $j = 1, 2, 3.$ Hence, 
\begin{align}\sum_{a = 1}^{N+3}R_{j a} [D_{i_{N + 3 - (a-1)}}] & =  \sum_{a = 1}^{N + 3} (Q^{-1} M)_{ja} [D_{i_{N + 3 - (a-1)}}] \\ & = 0\,,\end{align} for $j = 1, 2, 3$. Therefore, for each row of $R$, the divisor class corresponding to the first nonzero column of $R$ in that row can be expressed in terms of the following divisor classes. Hence, the divisor classes of the instantons in the $S$ of Algorithm \ref{alg:new_instanton_basis_determination} form a basis for $H_{4}(X_3)$ over $\mathbb{Q}.$ 

We now show that Algorithm \ref{alg:new_instanton_basis_determination} gives an equivalent result to Algorithm \ref{alg:old_instanton_basis_determination}. Let $S_1$ be the output of Algorithm \ref{alg:old_instanton_basis_determination}, and let $S_2$ be the output of Algorithm \ref{alg:new_instanton_basis_determination}. Suppose $I_j \in S_1 \setminus S_2.$ Since $I_j \in S_1$, $Q_{I_j}$ cannot be written as a linear combination of the $Q_{I_k}$ for $1 \leq k < j.$ However, since $I_j \not \in S_2$, there exists a $\mathbb{Q}$-linear combination of the divisor classes $[D_{i_1}], \ldots, [D_{i_{j - 1}}]$ which equals $[D_{i_j}].$ However, that means that $Q_{I_j}$ \textit{can} be written as a linear combination of the $Q_{I_k}$ for $1 \leq k < j,$ a contradiction. Hence, no such $I_j$ can exist. Thus, $S_1 \subseteq S_2$, but both $S_1$ and $S_2$ have $N$ elements, so we are done.

\subsection{Axion Lagrangian}

Taking $W_{\text{np}}$ to be given by 
\eqref{eq:wfullapp4}, we now evaluate the F-term potential.
Using
\begin{align}D_{T^a} W & = \frac{\partial}{\partial T^a} W + \left(\frac{\partial}{\partial T^a} \mathcal{K}\right)W \\ & = -\sum_a  e^{-S_a} \times  \frac{2 \pi}{c_2(D_a)} - \frac{W}{2\mathcal{V}} t_a\,,
\end{align} 
we have 
\begin{align}K^{a\bar{b}}(D_{T^a} W)(D_{\bar{T}^b}\bar{W}) & = 3|W|^2 + 8 \pi W_0 \text{Re}\left\{\sum_a   e^{-S_a} \frac{\tau^a}{c_2(D_a)}\right\} + \ldots
\end{align} 
where all  terms that are quadratic in $e^{-S_a}$ are lumped into the subleading terms.  
Using \eqref{eq:kother},
we then have 
\begin{align}
V_F & = -\sum_a \Lambda^4_a \cos\Bigl(2 \pi \theta^a\Bigr) + \text{subleading terms}\,,
\end{align} 
where
the instanton scales
are 
\begin{align}\Lambda_a^4 & = \frac{g_s^4}{128} \times \frac{8 \pi W_0}{\mathcal{V}^2} \frac{\tau^a}{c_2(D_a)} e^{-2 \pi  \tau^a/c_2(D_a)}\,.\end{align} 
Shifting the the total potential by a convenient constant (since we make no claim to address the cosmological constant problem here), 
the Lagrangian is given by 
\begin{align}
\mathcal{L} & \supset -\frac{1}{2} \sum_{a,b = 1}^N K_{ab}\,\partial_\mu \theta^a \partial^\mu \theta^b  - \sum_{a = 1}^N \Lambda_a^4 \Bigl(1 - \cos\bigl(2\pi \theta^a  \bigr)\Bigr) \nonumber \\ & - \sum_{\text{gauge group }\alpha}   \left(\tau^\alpha G_{\alpha}^{\; \mu\nu} G_{\alpha\mu\nu} + \theta^\alpha G_{\alpha}^{\mu\nu} \tilde{G}_{\alpha\mu\nu}\right)\,.
\end{align}  

\subsection{Change of Basis}

To compute axion-photon couplings, we need to canonicalize the kinetic term and write everything in terms of mass-eigenstate fields. Since $(K^{-1})^{ab}$ is positive and symmetric, we can perform a Cholesky decomposition, writing  
\begin{equation}(K^{-1})^{ab} = (L L^T)^{ab}\,,
\end{equation}  
where $L^a_{\; b}$ is a lower-diagonal $N \times N$ matrix with positive diagonal 
entries.\footnote{The map to the notation of \cite{glimmers} is as follows:
$L_{ab}^{\text{here}} = q_{ab}^{\text{there}}$,
where $q_{ab}^{\text{there}}$ is defined in equation (2.12) of \cite{glimmers}.} 
Then we can define fields $\phi^k$ with $k = 1, \ldots, N$ by setting 
\begin{equation}
\theta^a = \frac{1}{M_\text{pl}} L^a_{\; b} \phi^b\,,
\end{equation} 
so that writing the Lagrangian in terms of the $\phi^a$ canonicalizes the kinetic term. 
We can also redefine the gauge fields and coupling constants to pull a factor of 
$\frac{1}{2\sqrt{ \tau^\alpha}}$
out of each $G_{\alpha\mu\nu}$ term, so that we arrive at the Lagrangian 
\begin{align}
\mathcal{L} & \supset -\frac{1}{2} \sum_{a = 1}^N \partial_\mu \phi^a \partial^\mu \phi^a - \frac{1}{2}M_{ab} \phi^a \phi^b - \frac{1}{4}\sum_{\text{gauge group }\alpha} G_{\alpha}^{\; \mu\nu}G_{\alpha\mu\nu}  \nonumber \\ & - \frac{1}{4} \sum_{\text{gauge group }\alpha} \frac{1}{M_\text{pl}} \frac{1}{\tau^{\alpha}} \sum_{j, k = 1}^N L^\alpha_{\; a}\phi^a G_{\alpha}^{\;\mu\nu} \tilde{G}_{\alpha\mu\nu}\,,
\end{align} 
where the mass matrix $M_{ab}$ is given by 
\begin{equation}M_{ab}  := M_\text{pl}^2  (2 \pi)^2 \sum_{c = 1}^N (L^T)_a^{\; c} \Lambda^4_c (L)_{cb}\,. 
\end{equation} 
We now diagonalize the mass matrix,
\begin{equation}
    \text{diag}(m_a^2) = H^T M H\,,
\end{equation}
and write
\begin{equation}
\phi^c := H^c_{\; d} \varphi^d\,,
\end{equation} so that the $\varphi^a$ are mass eigenstates.
 
We arrive at the Lagrangian   
\begin{align}
\mathcal{L} & \supset -\frac{1}{2} \sum_{a = 1}^N \Bigl( \partial_\mu \varphi^a \partial^\mu \varphi^a  + m_a^2 (\varphi^a)^2\Bigr)   -  \frac{1}{4 M_\text{pl}}\sum_{\text{gauge groups }\alpha}  \mathscr{R}^{\alpha}_{\;\, b} \varphi^b G_{\alpha}^{\;\mu\nu} \tilde{G}_{\alpha\mu\nu}\,.
\end{align} 
where
\begin{equation}
    \mathscr{R}^{\alpha}_{\;\, b} :=  \frac{1}{   \tau^\alpha}\, L^{\alpha}_{\; c} H^c_{\; b}\,.~~\text{(no sum on $\alpha$)}
    \label{eq:curly_r_equation}
\end{equation}

\subsection{Axion-Photon Couplings}
 
If we define the axion-photon coupling $g_{a \gamma\gamma}$ of the $a$th axion to photons to be the term that appears in \begin{align}\mathcal{L} & \supset -\frac{1}{4} g_{a\gamma\gamma} \varphi^a F_{\mu\nu} \tilde{F}^{\mu\nu},\end{align} where $F_{\mu\nu}$ is the field strength of electromagnetism, then setting 
\begin{equation}
\alpha_{\text{EM}} = \frac{1}{\text{vol}(D_{\text{EM}})}\,,
\label{eq:fine_structure}\end{equation}
we have \begin{align}g_{a\gamma\gamma} = \frac{\alpha_{\text{EM}}}{M_\text{pl}} L^\text{EM}_{\; c} H^c_{\; a}\,,\label{eq:axion_photon_coupling_expression}\end{align} where we have used equation \eqref{eq:curly_r_equation} and set $\alpha \rightarrow \text{EM}$.\footnote{For reference we give the conversion from our result to that of \cite{glimmers}, which defines the mixing angle $\Theta_{ab}$ such that \begin{equation}2 \pi \frac{1}{M_\text{pl}} \sum_{b,c = 1}^N L_{ab} H^b_{\; c} \varphi^c = \sum_{b = 1}^N\Theta_{ab} \frac{\varphi^b}{f_b}\,,\end{equation} where $f_b$ is the decay constant of the $b$-th axion. Reexpressing $L$ and $H$ in terms of the mixing angle, and using $(n_\text{EM})_i L_{ic} = L^\text{EM}_c,$ we have \begin{equation}g_{a\gamma\gamma} = \frac{\alpha_\text{EM}}{2 \pi f_a} \sum_{i = 1}^N (n_\text{EM})_i \Theta_{ia} \times \frac{1}{M_\text{pl}}\,,\end{equation} reproducing equation (2.19) of \cite{glimmers}.} 

To compare to experiments, one takes the Pythagorean sum of the axion-photon couplings of axions below a cutoff mass $m_\text{cutoff}$. In addition, the $\alpha_\text{EM}$ quoted in equation \eqref{eq:fine_structure} is the coupling in the UV, so one should RG-flow it to its value of $\alpha_\text{EM} \simeq \frac{1}{137}$ in the IR. Hence, given a mass cutoff $m_\text{cutoff}$ we have effective couplings 
\begin{equation}\label{eq:appgeff}
g_\text{eff} = \frac{1}{137} \frac{1}{M_\text{pl}} \sqrt{\sum_a \bigl(L^{\text{EM}}_{~c} H^c_{\; a} \bigr)^2}\,,\end{equation}  
where the sum runs only over axions $a$ for which $m_a < m_\text{cutoff}$.
In the limit that $m_\text{cutoff} \rightarrow \infty,$ mass mixing induced by $M_{ab}$ is negligible, leading to
\begin{equation}
g_\text{eff, no cutoff} = \frac{1}{137} \frac{1}{M_\text{pl}} \sqrt{(K^{-1})_{\text{EM, EM}}}\,.
\label{eq:simple_geff}
\end{equation}

\section{Computation and Convexity}\label{sec:candc} 

Let $B_3$ be a smooth compact toric threefold base
for a Weierstrass model, and let $N = h^{1,1}(B_3)$. 
At some point we reach the situation where we are considering the subset $S$ of the K\"ahler moduli space of $B_3$ defined by the constraints that 
\begin{equation}\text{vol}(D_I) \geq a_I\,, \label{eq:alg_div_vol_constraint}\end{equation} for all $I = 1, \ldots, N + 3$ labelling the prime toric divisors of $X_3$ and for some $a_I > 0$; and \begin{equation}\text{vol}(C) \geq b_C\,,\label{eq:alg_curve_vol_constraint}\end{equation} for all toric curves $C$ and some curve-dependent $b_C \geq 0.$ For the most part, we take $S$ to be the dual Coxeter stretched K\"ahler cone defined in \eqref{eq:dcskc_def}, i.e.~we take $b_C = 1$ and $a_I = c_2(D_I)$, but here we work in slightly more generality, just demanding that the $a_I$ are positive and the $b_C$ are non-negative. 

Within $S$, we then desire to find a reference K\"ahler form $J$ for use in calculating quantities related to axion physics. If we let $\omega^1, \ldots, \omega^{N + 3}$ be the $(1, 1)$-forms Poincar\'e dual to the divisor classes $[D_1], \ldots, [D_{N + 3}],$ we can expand the K\"ahler form $J$ as  \begin{equation}J = \sum_{I = 1}^{N + 3} t_I \omega^I\,. \label{eq:kahler_totdcskc_def}\end{equation} A good reference $J$ would be the form in $S$ that minimizes the overall volume $\mathcal{V}$ of $B_3.$ However, finding such a point would involve solving a cubic programming problem for the $t_I$. This is too hard at large $N.$ Instead, we settle for a point $t_\star = (t_1, \ldots, t_{N + 3}) \in S$ of minimal $\ell^1$   or $\ell^2$ norm, and take this as a proxy for the $\mathcal{V}$-minimizing point.\footnote{One might observe that \eqref{eq:kahler_totdcskc_def} is over-parameterized, since there are three linear relations among the $\omega^I.$ However, if one eliminates three of the $\omega^I,$ then the point in K\"ahler moduli space one finds by minimizing a norm of the remaining $t_I$ may be basis-dependent. }

Provided that $S$ is convex, one can then find  an arbitrarily good approximation to $t_\star$, assuming that one is in possession of a program that can solve either linear programs in the $\ell^1$ case, or linearly constrained programs with quadratic objectives in the $\ell^2$ case. We describe the algorithm for computing approximations to $t_\star$ in \S\ref{sec:tstar}, and prove that $S$ is convex in \S\ref{sec:convex}.

\subsection{Finding a point $t_\star$ of minimal norm}\label{sec:tstar}
Define the triple-intersection numbers $\kappa^{IJK}$ for $I, J, K = 1, \ldots, N + 3$ by 
\begin{equation}\kappa^{IJK} := \int_{B_3} \omega^I \wedge \omega^J \wedge \omega^K\,.\end{equation} For later notational convenience, for each $I = 1, \ldots, N + 3$, let $M_I \in \mathbb{R}^{N + 3} \times \mathbb{R}^{N + 3}$ be the matrix with entries 
\begin{equation}(M_I)_{JK} = \kappa^{IJK}\,. \label{eq:matrix_m_definition}\end{equation} Then the task is, for $p = 1$ or $2,$ to solve the mathematical program 
\begin{align} 
\min ||t||_p \\ \sum_{K = 1}^{N+3}\kappa^{IJK} t_K & \geq b_{IJ} & \text{ for all } I, J = 1, \ldots, N+3 \label{eq:alg_linear_constraint} \\ \frac{1}{2} t^T M_I t & \geq a_I & \text{for all }I = 1, \ldots, N+3\,, \label{eq:alg_quadratic_constraint}\\ t & \in \mathbb{R}^{N+3}\,, & & \label{eq:t_in_r}
\end{align} 
where in equation \eqref{eq:alg_linear_constraint}, $t_K$ is the $K$-th entry of $t.$ 

Using an appropriate backend such as Gurobi \cite{gurobi},
it is easy to solve, for $p = 1$ or 2, the program  
\begin{align} \label{eq:minp}
\min ||x||_p & &\\ 
M x & \geq c \\ x 
& \in \mathbb{R}^j\,, \label{eq:minp3}
\end{align} for some matrix $M \in \mathbb{R}^{i \times j},$ and $c \in \mathbb{R}^i$.   
If one does not enforce the constraints in equation \eqref{eq:alg_quadratic_constraint}, then the problem we desire to solve is in the format of the easily solvable problem \eqref{eq:minp}-\eqref{eq:minp3}. In particular, we would take $j = N + 3$, set $t = x$, set $i = (N+3)^2$, take $M$ to be  
$(\kappa^{11K}, \kappa^{12 K}, \ldots)^T,$
and take 
$c = (b_{11}, b_{12}, \ldots)^T.$ 
Since $S$ is convex, one can make arbitrarily good approximations to $S$ by taking linear outer approximations to the quadratic constraints. One solves the problem specified by equations \eqref{eq:alg_linear_constraint} and \eqref{eq:t_in_r}, dilates the obtained point until it satisfies equation \eqref{eq:alg_quadratic_constraint}, and appends the tangent hyperplane at that dilated point to the set of constraints to be solved in the next iteration. Proceeding in this way, one can, given any $\epsilon > 0$ and enough iterations, solve for a point of minimal $\ell^p$ norm that exactly satisfies equations \eqref{eq:alg_linear_constraint} and \eqref{eq:t_in_r}, and satisfies equation \eqref{eq:alg_quadratic_constraint} with $a_I$ replaced with $a_I(1 - \epsilon).$ In our work, we take the tolerance $\epsilon = 0.01.$

\subsection{Proof that $S$ is convex}\label{sec:convex}

It is not immediately obvious that $S$ is convex. Note that the region $S'$ defined by just $\text{vol}(D_I) \geq a_I$ is not convex. By finding a K\"ahler form and dilating appropriately, one obtains an element of $S'$, so $S'$ is nonempty. However, since the $a_I > 0$, $J = 0$ is not in $S'$. That being said, since the divisor volumes are invariant under $J \mapsto -J$, convexity of $S'$ would imply that $J = 0$ would in fact be in $S'$, a contradiction. Hence, the curve volume constraints must play a role in ensuring that $S$, which is $S'$ additionally constrained by equation \eqref{eq:alg_curve_vol_constraint}, is convex.

To prove that $S$ is convex, we first prove two lemmas. In the proof of the final result, we will show that any convex combination of K\"ahler forms in $S$ remains in $S$. To do this, we need a lower bound on the volume of a given divisor as measured by the combination of K\"ahler forms, in terms of its volume as computed by each of the original K\"ahler forms. The first lemma provides information about the eigenvalues of the $M_I$ defined in equation \eqref{eq:matrix_m_definition}, showing that the $M_I$ fulfill the first part of the hypotheses of the second lemma, which provides the desired bound. 

\newtheorem{lemma}{Lemma}
\begin{lemma}\label{lemma:eig}
For arbitrary $I$ between 1 and $N + 3$, let $M_I$ be the $(N + 3) \times (N + 3) $-dimensional matrix given by equation \eqref{eq:matrix_m_definition}. Then $M_I$ has exactly one eigenvector with positive eigenvalue. 
\end{lemma}

\textit{Proof:} Note that $M_I$ is the tensor product of a matrix of zeros with the intersection numbers on the prime toric divisor $D_I \subsetneq B_3$. Hence, it is enough to show that the intersection numbers of $D_I$ have exactly one eigenvector with positive eigenvalue. 

The Hodge Index Theorem says that if $V$ is a nonsingular projective surface,\footnote{We expect that the results of this appendix could be generalized to threefolds $B_3$ that are not toric, because the essential requirement in what follows is that the divisors in question are nonsingular and projective.} then the intersection numbers on $V$ have exactly one eigenvector with positive eigenvalue \cite{Voisin_2002}.\footnote{Thanks to Elijah Sheridan for pointing this theorem out to us.} 
Moreover, every complete nonsingular surface is projective
(see e.g.~\cite{hartshorne1977algebraic}.)
Hence, it is enough to show that $D_I$ is nonsingular and complete. If $D_I$ were singular, then since it is a toric variety itself, there would exist a cone in the fan for $D_I$ whose minimal generators did not form a part of a $\mathbb{Z}$-basis for $\mathbb{Z}^2.$ However, then the corresponding cone in the fan for $B_3$ would not have minimal generators forming part of a $\mathbb{Z}$-basis for $\mathbb{Z}^3$, contradicting the smoothness of $X_3$.\footnote{Because $B_3$ is smooth, given any three minimal ray generators in the fan of $B_3$, one can find a $\text{GL}(n, \mathbb{Z})$ rotation of the fan that sets the minimal generators to the standard basis vectors. Hence, in the fan of $D_I$, all cones must be smooth. } Therefore, $D_I$ is smooth. It remains to show that $D_I$ is complete. By theorem 3.4.6 of \cite{CLS}, a toric variety is compact in the classical topology if and only if it is complete, which is true if and only if its fan is complete. If the fan of $D_I$ were not complete, then the fan of $B_3$ would not be complete, contradicting the compactness of $B_3.$ Thus $D_I$ is complete, and so projective. Hence, the conclusion of the Hodge Index Theorem applies to $D_I$, and so $M_I$ has exactly one eigenvector with positive eigenvalue. $\square$

In the next lemma, we establish the lower bound we will use in proving that $S$ is convex. This is essentially a technical lemma, but note that, by Lemma \ref{lemma:eig}, the matrices $M_I$ each satisfy the hypothesis exclusively concerning the $M \in \mathbb{R}^{n,n}.$ 

\begin{lemma}\label{lemma:eig2}
Suppose $M \in \mathbb{R}^{n,n}$ is a symmetric real-valued matrix with exactly one positive eigenvalue; and suppose the components of $t_1, t_2 \in \mathbb{R}^n$ along the associated eigenvector have the same sign. Then if $t_1^T M t_1$, and $t_2^T M t_2$ are both positive, we have that 
\begin{align}(t_1 + t_2)^T M (t_1 + t_2) \geq t_1^T M t_1 + t_2^T M t_2 + 2 \sqrt{t_1^T M t_1} \sqrt{t_2^T M t_2}\,.\end{align} 
\end{lemma}

\textit{Proof:} Let $v_1, v_{2}, \ldots, v_k, v_{k+1}, \ldots, v_n$ be the eigenvectors of $M$ with eigenvalues $\lambda_1, \ldots, \lambda_n$. Without loss of generality, we can choose $v_1$ to have positive eigenvalue, $v_2, \ldots, v_k$ to have negative eigenvalue, and $v_{k+1}, \ldots, v_n$ to have zero eigenvalue. Since this is a basis, we can write 
\begin{align}t_1 = a_1 v_1 + \cdots + a_{n} v_n\,,\end{align} and
\begin{align}t_2 = b_1 v_1 + \cdots + b_{n} v_n\,.\end{align} Then 
\begin{align}(t_1 + t_2)^T M (t_1 + t_2) & = \lambda_1 (a_1 + b_1)^2 + \lambda_{i+1} (a_{i+1} + b_{i+1})^2 + \cdots + \lambda_k (a_k + b_k)^2 \\ & = |\lambda_1| (a_1 + b_1)^2 - |\lambda_{i+1}| (a_{i+1} + b_{i+1})^2 - \cdots - |\lambda_k| (a_k + b_k)^2 \\ & =  (\tilde{a}_1 + \tilde{b}_1)^2 - (\tilde{a}_{i+1} + \tilde{b}_{i+1})^2 - \cdots - (\tilde{a}_k + \tilde{b}_k)^2 \\ & = t_1^T M t_1 + t_2^T M t_2 + 2 \left(\tilde{a}_1 \tilde{b}_1 - \tilde{a}_2 \tilde{b}_2 - \cdots \tilde{a}_k \tilde{b}_k\right)\,,\end{align} where $\tilde{a}_i = \sqrt{|\lambda_i|} a_i$ and the $b_i$ are defined similarly. By two applications of Cauchy-Schwarz we can rewrite 
\begin{align}\tilde{a}_1 \tilde{b}_1  - \tilde{a}_{i+1} \tilde{b}_{i+1} - \cdots - \tilde{a}_k \tilde{b}_k & \geq \tilde{a}_1 \tilde{b}_1  - |\tilde{a}_{i+1} \tilde{b}_{i+1}| - \cdots - |\tilde{a}_k \tilde{b}_k| \\& \geq \tilde{a}_1 \tilde{b}_1 - \sqrt{\tilde{a}_{i+1}^2 + \cdots + \tilde{a}_k^2}\sqrt{\tilde{b}_{i+1}^2 + \cdots + \tilde{b}_k^2} \\ & =  \tilde{a}_1 \tilde{b}_1 - \sqrt{\tilde{a}_1^2 - t_1^T M t_1}\sqrt{\tilde{b}_1^2 - t_2^T M t_2} \\ & = \sqrt{t_1^T M t_1}\sqrt{t_2^T M t_2} + \tilde{a}_1 \tilde{b}_1\\\nonumber &   - \bigg(\sqrt{\tilde{a}_1^2 - t_1^T M t_1}\sqrt{\tilde{b}_1^2  - t_2^T M t_2} + \sqrt{t_1^T M t_1}\sqrt{t_2^T M t_2}\bigg)  \\ & \geq \sqrt{t_1^T M t_1}\sqrt{t_2^T M t_2} + \tilde{a}_1 \tilde{b}_1  - \sqrt{\tilde{a}_1^2} \sqrt{\tilde{b}_1^2} \\ & = \sqrt{t_1^T M t_1}\sqrt{t_2^T M t_2}\,,\end{align} as required. Note that the last step follows by the hypothesis that $\tilde{a}_1$ and $\tilde{b}_1$ have the same sign. Otherwise, the last two terms on the right hand side do not cancel. $\square$

With Lemma \ref{lemma:eig} and Lemma \ref{lemma:eig2} in hand, we can prove that $S$ is convex.
 
\begin{proposition}\label{prop:convex}
The subset $S$ of K\"ahler moduli space satisfying inequalities \eqref{eq:alg_div_vol_constraint} and \eqref{eq:alg_curve_vol_constraint} is convex. In particular, the dual Coxeter stretched K\"ahler cone  defined in \eqref{eq:dcskc_def} is convex.
\end{proposition}

\textit{Proof:} Suppose $t_1, t_2 \in \mathbb{R}^{N+3}$ satisfy \eqref{eq:alg_linear_constraint} and \eqref{eq:alg_quadratic_constraint}. Then we are done if we can show that 
\begin{equation}
t(\theta) := (1 - \theta) t_1 + \theta t_2 \, ,
\end{equation} 
is contained in $S$ for all $\theta \in [0,1].$ We do this by first showing that the hypotheses of Lemma 2 hold for $t_1$ and $t_2$ with each of the matrices $M_I$, and then we use the conclusion of Lemma 2 to establish the result. 
    
Let $t^\star$ be a point in the $1$-stretched K\"ahler cone, i.e. the region of K\"ahler moduli space where constraint \eqref{eq:alg_linear_constraint} holds with $b_{IJ}$ set to $1$ for all $I, J$. Then let $s \in S$ be arbitrary and set $s(\omega) = (1 - \omega) t^\star + \omega s$ for $\omega \in [0, 1]$. Note that $s(\omega)$ is contained in the interior of the K\"ahler cone for $\omega > 0$, although if $s$ is on the boundary of the K\"ahler cone, then $s(\omega)$ is not in the interior for $\omega = 0$. Let $I$ be an arbitrary integer between 1 and $N + 3$, and let $v_I$ be the unique eigenvector of $M_I$ with positive eigenvalue. Then consider $v_I \cdot s(\omega)$. This is a linear function of $\omega,$ and in particular can never be zero, because otherwise the divisor corresponding to the matrix $M_I$, i.e.~$D_I$, would have at most zero volume, which is impossible in the interior of the K\"ahler cone, and is also impossible at $s$, since $a_I > 0$. In other words, $v_I \cdot s(\omega)$ has the same sign for all $\omega$. Hence, since the K\"ahler cone is convex, if $s, s' \in S$ then one can connect $s$ to $t^\star$ with a line, and then connect $t^\star$ to $s'$ with a line, thus allowing one to conclude $\text{sign}(v_I \cdot s) = \text{sign}(v_I \cdot s')$ for all $I.$

Now let $1 \leq I \leq N + 3$ and $\theta \in [0,1]$ be arbitrary. Note that $M_I$ is symmetric with real entries and has only one positive eigenvalue. Moreover, the argument of the previous paragraph shows that the components of $t_1$ and $t_2$ along that eigenvector have the same sign. Also, since $t_1$ and $t_2$ satisfy constraint \eqref{eq:alg_quadratic_constraint}, both $t_1^T M t$ and $t_2^T M t_2$ are positive. Hence, by Lemma \ref{lemma:eig2}, we have 
\begin{align}
\frac{1}{2}t(\theta)^T M_I t(\theta) & \geq \frac{1}{2} \times \left\{(1 - \theta)^2 t_1 M_I t_1 + \theta^2 t_2 M_I t_2 + 2 \theta (1 - \theta) \sqrt{t_1^T M_I t_1} \sqrt{t_2^T M_I t_2} \right\} \\ & \geq \left[(1 - \theta)^2 + \theta^2 + 2\theta (1 - \theta) \right] a_I\\ & = a_I\,.
\end{align} 
The above shows that $t(\theta)$ satisfies \eqref{eq:alg_quadratic_constraint} for all $\theta \in [0, 1]$. It satisfies \eqref{eq:alg_linear_constraint} by linearity. Hence, $t(\theta) \in S$ for all $\theta \in [0, 1].$ $\square$

Note that the proof hinges on the equality 
\begin{equation}\text{sign}(v_I \cdot t_1) = \text{sign}(v_I \cdot t_2)\,,\end{equation} for $t_1, t_2 \in S.$ In the absence of the linear constraints, this equality no longer holds, because both $t_1$ and $-t_1$ satisfy the quadratic constraints \eqref{eq:alg_quadratic_constraint}.      


\end{appendix} 
\bibliographystyle{utphys}
\bibliography{ref}

\end{document}